\documentclass[12pt]{article}
\usepackage{arxiv}
\usepackage{amsmath}
\usepackage{amssymb}
\usepackage{float}

\title{A Bayesian workflow for securitizing casualty insurance risk}
\author[1]{Nathaniel Haines\thanks{nathaniel.haines@ledgerinvesting.com}}
\author[1]{Conor Goold\thanks{conor@ledgerinvesting.com}}
\author[1]{J. Mark Shoun\thanks{mark@ledgerinvesting.com}}
\affil[1]{Ledger Investing, Inc.}

\begin{document}
\maketitle

\begin{abstract}
    Casualty insurance-linked securities (ILS) are appealing to investors because the 
underlying insurance claims, which are directly related to resulting security 
performance, are uncorrelated with most other asset classes. Conversely, casualty ILS 
are appealing to insurers as an efficient capital management tool. However, 
securitizing casualty insurance risk is non-trivial, as it requires forecasting loss 
ratios for pools of insurance policies that have not yet been written, in addition to 
estimating how the underlying losses will develop over time within future accident years. 
In this paper, we lay out a Bayesian workflow that tackles these complexities by using: 
(1) theoretically informed time-series and state-space models to capture how loss ratios 
develop and change over time; (2) historic industry data to inform prior distributions of 
models fit to individual programs; (3) stacking to combine loss ratio predictions from 
candidate models, and (4) both prior predictive simulations and simulation-based 
calibration to aid model specification. Using historic Schedule P filings, we then show 
how our proposed Bayesian workflow can be used to assess and compare models across a 
variety of key model performance metrics evaluated on future accident year losses.
\end{abstract}


\section{Introduction}
Insurance-linked securities (ILS) are financial instruments that enable insurers to 
transfer risk to the capital markets. Traditionally used in the context of 
catastrophe insurance risk \citep[e.g.][]{ils2022, lane2024}, ILS have 
arisen as effective tools for capital management, providing insurers with sources of 
capacity from outside the traditional reinsurance industry and investors with access to 
a novel asset class that is relatively uncorrelated with other assets in the 
marketplace \citep{jaeger2010}. 

The insurance market is often described as cyclical, where profits grow and fall 
periodically as underwriters observe and respond to both micro- and macro-economic 
factors that impact program performance \citep[e.g.][]{berger1988}. Although the 
existence of true, predictable periodicity in insurance underwriting is debated, periods 
where premiums become higher or lower relative to losses are readily observed 
\citep[e.g.][]{boyer2012}. In recent years, ``hard'' market conditions have lead to a 
significant increase in attention given to the problem of securitizing casualty insurance 
risks. The reason is straightforward -- individual casualty insurers, and subsequently the 
casualty insurance market as a whole, can only hold a limited amount of risk. As capacity 
for risk reaches the limit, the ability to write new risk and pursue new opportunities 
diminishes, resulting in stagnated growth. In such conditions, casualty insurers need to 
source capital efficiently so that they can take advantage of opportunities while not 
diluting existing shareholders through other capital management strategies. ILS provides 
exactly this flexibility, allowing insurers to release capital otherwise locked 
in long-tailed reserves and then redeploy it in other areas \citep[see][]{canabarro2000}. 
This ability to access external capital allows insurers to compete more effectively, 
control more business than their balance sheets would typically support, and signal the 
quality of their underwriting through favorable terms from ILS investors.

Despite the promise of casualty ILS, there are significant technical challenges 
involved in securitizing the underlying casualty insurnace risk. First and foremost, 
casualty insurance is notoriously ``long-tailed'', meaning that it can take years or 
even decades for claims to be settled depending on the specific line of business. For 
example, for worker's compensation polices written in 2024, we may not known the losses 
incurred by associated claims until 2034, 2044, or even later depending on the nature 
of the risk \citep{tailfactors2013}. This long-tailed behavior means that 
the true losses associated with a casualty insurance portfolio are often not known for years down the 
road, presenting a significant challenge for pricing casualty ILS. This uncertainty is further 
compounded in cases where the insurance program in question has sparingly little historic data 
available to use for informed decision-making, as is often the case with new products seeking 
reinsurance or ILS opportunities.

Due to its flexibility and the ease at which it can accommodate external information, 
the Bayesian framework is particularly well-suited to address the complexity and 
uncertainty associated with modeling casualty insurance risk. Bayesian models are
used across a diverse range of industries to address exactly these issues, including 
in pharmaceuticals \citep{lesaffre2020}, energy \citep{adedipe2020}, marketing 
\citep{wang2017}, economics and finance \citep{martin2023}, and insurance loss reserving 
more generally \citep{meyers2015}. Perhaps most conceptually relevant to our work, 
Bayesian models have also been proposed for use in valuation of catastrophe bonds 
\citep{domfeh2023}, a form of catastrophe ILS. However, catastrophe bonds have a 
decades-long history of use \citep[see][]{lane2024}, and the underlying risk models are 
generally well-agreed upon \citep[see][]{mitchell2017}. By contrast, casualty ILS has no 
standardized modeling framework, in part because it is a relatively novel financial 
instrument.

Here, we present a Bayesian workflow for modeling the long-tailed, uncertain nature of 
casualty insurance risk. We use both simulations and historic real-world data to show how 
a combination of theory-informed models, carefully-specified priors, and data-driven 
model averaging and selection can be used to accurately forecast and quantify uncertainty 
in how losses develop over time across multiple lines of business, thus making casualty 
ILS feasible. Our aim in presenting this workflow is to catalyze further research on 
casualty ILS, which we believe is essential for enhancing its practical use and promoting 
wider adoption by both insurers and capital providers alike. 

Note that we assume basic familiarity with Bayesian analysis throughout the following 
sections. For readers not yet familiar with Bayesian analysis, we recommend 
\cite{mcelreath2018} and \cite{bda2013} for in-depth coverage and 
\cite{vandeschoot2021} for a concise primer.

\section{What is a Bayesian modeling workflow?}
\label{section:bayes-workflow}
A ``Bayesian workflow'' formalizes the iterative, and often non-linear, 
process of development, fitting, validation, 
comparison, and selection or averaging of Bayesian models
\citep{gelman2020,gabry2019}. Particular workflows for Bayesian
modeling can now be found for fields such as cognitive science
\citep{schad2021} and epidemiology \citep{grinsztajn2021,bouman2024},
but there has been no discussion of what a Bayesian workflow 
should look like for actuarial science, let alone for casualty ILS 
modeling in particular \citep[although Bayesian methods are gaining popularity, see][]{dealba2002,meyers2015}.
This is particularly important for actuarial science
due to the multi-stage analyses that are frequently
conducted, from loss development to forecasting of future
ultimate loss ratios. Enumerating a Bayesian workflow
is necessary to enhance robustness and reproducibility
of actuarial analyses that use Bayesian methods, including
analyses needed for casualty ILS.

Below, we provide an introduction to the main components of a 
Bayesian workflow that are relevant to our specific workflow for
modeling casualty insurance risks presented in a later section. 
We use a simplified Bayesian chain-ladder loss development model 
as a running example here, although we provide fuller illustration in 
Section \ref{section:ledger-workflow}.

\subsection{Model development and fitting}
\label{section:chain-ladder-intro}
Model development typically starts from the selection
of a base model, which might be an existing model
in the literature, or a newly developed model.
In either case, it is important to clearly reify the
the structure of relationships between 
observed variables and model parameters
mathematically, in code, and/or graphically \citep{kruschke2021}.
Commonly, Bayesian
models are at least partly
generative \citep{gelman2020}, 
specifying a joint probability distribution
over the observed data and parameters, and
allowing for forward simulation
of new data from the data generating process.

It is helpful to consider a concrete example when describing 
model development. For loss development models in insurance, 
claims across most lines of business often take many months, 
or even years, to settle. For example, if a policy holder is 
in a car accident in November of accident year 2024, it may 
take a few months for their car to be repaired in an auto shop. 
Further, if they experienced bodily injury as a result of the 
accident, medical claims could persist for years into the future. 
When viewed in aggregate across policies (i.e. at the ``program'' level), 
this ``loss development'' means that the losses incurred by a 
pool of policies in a given accident year are not fully known 
until years into the future. Depending on the underlying risk, 
it could even take decades for losses within a given accident 
year to reach their ``ultimate'' state. 

Loss development modeling is the practice of using historic 
patterns of loss development to predict ultimate losses for 
all accident years within a pool of policies. Data are typically 
organized into a triangular matrix of experience periods 
by development periods. For our purposes, the experience period 
is the year an accident occurs, and the development period is the 
number of years (or lags) since the accident year. Values in the 
triangle can be cumulative or incremental losses that indicate the observed loss for 
each accident year as of each development lag (we focus on cumulative losses here). 
Generally, we expect that losses within each accidenty year will asymptote to some 
ultimate loss as development lag increases. Some lines of business are relatively 
quick to develop (e.g., private passenger auto), whereas others can 
often take many years (e.g., worker's compensation).

One may start modeling such loss development using a 
Bayesian variation of the
chain-ladder method \citep{mack1993, englandverrall2002} as the base.
The joint probability distribution of
the losses, $y$, and parameters then might
be represented as $p(y, \alpha, \sigma)$,
where $\alpha$ is an $M - 1$ vector of link ratios, one
for each development lag save the first development
period, and $\sigma$
is the residual standard deviation. The joint density
factors into the likelihood distribution, $p(y \mid \alpha, \sigma)$,
and (independent) prior distributions 
$p(\alpha, \sigma) = p(\alpha) p(\sigma)$.
The likelihood distribution for the $i$th accident year
and $j$th development period would be a positive-bound
probability density function, such as the lognormal distribution,
with a mean determined by a multiplicative, autoregressive lag-1
formulation typical of chain-ladder models, 
i.e. $p(y_{ij} \mid y_{ij - 1}, \alpha_{j - 1}, \sigma) = 
\mathrm{Lognormal}(y_{ij} \mid \log(\alpha_{j - 1} y_{ij - 1}), \sigma)$.

The choice of prior distribution is, historically, one of the most
contended aspects of Bayesian modeling, and one must justify
why priors were selected \citep{winkler1967,kruschke2021,mikkola2023}. 
We support, as a starting
point, the use of \textit{weakly-informative} prior distributions that are
sufficiently diffuse to allow a broad range of parameter values
without reducing model performance (e.g. resulting in underflow
or overflow of numerical computations) or realism. 
If independent expert knowledge is available, prior elicitation
from multiple domain experts may be used, or from previous
data if independent datasets are available \citep{falconer2022,mikkola2023}.
Nonetheless, weakly informative prior distributions are still
a useful starting point for model development because
they appeal to the principles of starting simply 
and failing fast \citep{gelman2020}, from which more informative
priors can be developed.
It is often wise to scale the data, or choose model parameterisations, 
in such a way that a standard normal distributions, or other generic
prior distributions, would
satisfy the definition of weakly informative. For instance,
for the chain-ladder model above, a standard normal prior
on the log-scale link ratios, $\log \alpha$,
would imply a median link ratio of 1.0 on the 
lognormal scale, with a mean of 1.7 and a standard deviation
of 2.2, which is suitably realistic yet diffuse
enough to capture a range of average loss development dynamics.
For $\sigma$, the residual standard deviation, 
we might first choose to scale the losses by its standard deviation,
or some large value,
to ensure that a standard normal prior distribution
on $\log \sigma$ is also appropriate. These priors are easily
generalizable to different datasets if the same
data scaling is applied.

Together, a variant of the chain-ladder model can be written 
mathematically as:

\begin{align}
	\begin{split}
	y_{ij} &\sim \mathrm{Lognormal}(\mu_{ij}, \sigma)\\
	\mu_{ij} &= \log(\alpha_{j - 1} y_{ij - 1})\\
	\log \alpha &\sim \mathrm{Normal}(0, 1)\\
	\log \sigma &\sim \mathrm{Normal}(0, 1)
	\end{split}
\end{align}

A useful exploratory method of checking model assumptions
during model development is prior predictive checks 
\citep{gabry2019,kruschke2021,gelman2020}. Prior
predictive checks simulate data from the full generative
model without fitting the model, which are samples
from the prior predictive distribution,
allowing modelers to compare the simulated data
to \textit{a priori} expectations and expert knowledge.
Prior predictive checks are particularly useful when working 
with complex models where the implications of a set of parameters 
on resulting losses is unintuitive or otherwise difficult to 
reason about, as is the case with most models that we use regularly. 
For example, wide priors on the chain-ladder link ratios ($\alpha$) 
can create an ``explosion'' or ``crash'' in loss predictions across 
development if they are too high or too low, leading to numerical 
overflow and underflow, respectively. Ideally, priors can be selected 
to ensure that losses roughly grow and eventually plateau in each accident 
period, leaving enough room for variable types of dynamics, but not allowing
for degenerate model predictions. We leave specific examples of prior 
predictive checks for Section \ref{section:prior-posterior-predictions}.

In this paper, we fit all Bayesian models 
using Stan \citep{stan2017}, via
the Python interface \texttt{CmdStanPy} \citep{cmdstanpy2024} and
command line interface \texttt{CmdStan} \citep{cmdstan2024}.
Stan uses Hamiltonion Monte Carlo, a variant of Markov chain
Monte Carlo, to sample from the posterior distribution,
and returns a number of diagnostic criteria for modelers
to inspect the convergence of their Markov chains.
We do not go into the details of convergence
diagnostics for Markov chain Monte Carlo here,
as these have been covered in detail elsewhere
\citep[e.g.][]{bda2013}.
Our Bayesian workflow is agnostic to software or
implementation choice, and so our discussion below
is not particular to Stan. 

\subsection{Model validation}

Validation of Bayesian models ideally starts before data is collected.
While prior predictive checks are one type of model validation,
a more principled approach fits the model to multiple sets of
data simulated from the prior predictive distribution
to inspect the accuracy and calibration of parameter recovery.
The now dominant paradigm for the latter task is 
\textit{simulation-based calibration} \citep{talts2018,modrak2023}, 
which is motivated by
the \textit{self-consistency} property of Bayesian models. Put
simply, fitting the model to many instances of data simulated
from the prior predictive distribution, and averaging across 
each resulting posterior distribution, 
should return the prior distributions
for each parameter, or other quantity of interest. 
Taking the first link ratio in the Bayesian
chain-ladder example, $\alpha_{1}$, if the model is working
as intended, we would expect the posterior means of
$\log \alpha_{1}^{\prime (m)}$,
where the superscript indicates a posterior distribution (the prime symbol) from
the $m$th prior predictive sample, 
to match the standard normal prior distribution,
$p(\log \alpha) = \mathrm{Normal}(0, 1)$.
It is important to also inspect the distribution of
test quantities that fold in the complete data space,
such as the joint log likelihood \citep{modrak2023}.
Typically, the posteriors are summarized not by their
means but using rank order statistics \citep{casella2002},
$R(\theta) = \sum_{s=1}^{S} \theta > \theta^{\prime (s)}$,
where $\theta$ represents any parameter or generated
quantity from the prior predictive distribution, and
$\theta^{\prime (s)}$ is the $s$th sample from the estimated
posterior distribution. If the model is working
as intended, then a histogram of $R(\theta)$ should be approximately
uniformly distributed, for which graphical and numerical
tests of uniformity can be used to confirm. 
Departures from uniformity indicate biases in parameter
estimation (skewed histograms), estimates that are
too certain (U-shaped histograms),
or estimates that are too uncertain (inverted U-shaped histograms).

Simulation-based calibration additionally provides the modeler
access to a large number model fits, from which
other useful model validation quantities can
be inspected.
For instance, the proportion of time a parameter $\theta$
falls within some interval of the estimated posterior distribution
$\theta^{\prime}$ across simulated datasets
informs the modeler about parameter calibration,
and the percentile of the real observations on the posterior
predictive distribution informs us about the calibration
of the model forecasts \citep{gneiting2007calibration}.
Moreover, convergence diagnostics for the Markov chain Monte
Carlo samplers can be aggregated and any systematic computational
problems identified.

In addition to simulation-based calibration, model validation
can make use of posterior predictive checks. Like 
prior predictive checks, posterior predictive checks 
can be both numerical or purely graphical inspections
of model fit to the data \citep{guttman1967,rubin1984,gelman1996}, 
most usefully the actual data, or types of data, that the model will be 
fit to in its real-world applications. Posterior predictive checks 
allow for us to determine if a model is able to capture qualitiative 
features in the observed data that align with expert opinion or intuition. 
For example, loss development models should be specified such that the uncertainty 
in predicted losses levels off as development lag increases. Leveling off is 
expected due to the fundamentals of how insurance claims are paid off over time 
until there are none left to pay, leading to asymptotic behavior of 
losses across development lags within accident years. If posterior 
predictions from the chain-ladder model fitted to real-world data 
show that losses grow without bound across development, it is likely 
that the model is improperly specified with respect to the data-generating 
process in some way. 

An additional benefit of posterior predictive checks is that they can be used to 
compute other quantities of interest, including both in- and out-of-sample fit 
statistics that can be used for model comparison. In our work, we often inspect 
calibration of the posterior predictive distribution by calculating the proportion 
of true values that fall within some interval width of the posterior samples,
or compute some quantity of distance of the prediction from
the true values \citep[see e.g.][for futher examples]{rubin1984,
gneiting2007calibration,gelman2020}. Such metrics can be used to gain a high-level
understanding of model performance, including if the model makes predictions that
are biased or over/under-confident in some way. 

\subsection{Model comparison, selection and averaging}
\label{section:model-compare-intro}

The final main component of the Bayesian workflow we cover here
is how to approach the comparison of competing models,
and the eventual selection or averaging of competing models.
Model comparison, selection, and averaging is an expansive
topic, from the choice of quantites to use to score
models \citep[e.g.][]{gneiting2007,vehtari2012,piironen2017},
to the method of model averaging 
\citep[e.g.][]{carlinchib1995,hoeting1999,yao2018,yao2022}.
Below, we follow modern Bayesian model comparisons by
using the expected log pointwise predictive
density (ELPD) \citep{vehtari2017} to score models, 
which is based on the logarithmic 
score \citep{good1952} 
and defined as:

\begin{align}
	\begin{split}
		\ELPD &= \log \int p(\tilde{y} \mid y) d \theta\\
		&= \log \int p(\tilde{y} \mid \theta) p(\theta \mid y) d \theta\\
		&\approx \sum_{i=1}^{N} \frac{1}{S} \log p(\tilde{y}_i \mid \theta^{s})
		p(\theta^{s} \mid y)\\
		&\approx \sum_{i=1}^{N} \ELPD_i
	\end{split}
\end{align}

where $\tilde{y}$ is external, out-of-sample data not used during model fitting,
and the approximation in the last line reflects the Monte Carlo approximation
of ELPD from posterior samples. Conceptually, ELPD is a measure of the 
height of the posterior predictive density at the true value for each datapoint, 
aggregated across all out-of-sample datapoints. We focus on out-of-sample predictive
ablity when comparing models, rather than in-sample measures of model
fit. The latter is useful for posterior predictive checks,
and model development cycles, whereas the former matches the goals
of predictive inference, and securitizing casualty insurance risks,
more closely. When external data to score models are not available,
approximate methods may be employed, namely cross validation,
such as approximate leave-one-out cross-validation via Pareto
smoothed importance sampling \citep{vehtari2017}.
For two models, $a$ and $b$, the difference between ELPDs
is often of interest, 
$\ELPD_{\mathrm{diff}} = \ELPD_{a} - \ELPD_{b}$,
and potential measures of uncertainty 
such as the standard error of the differences
\citep{vehtari2017,sivula2023}.
Other scoring rules are of course possible,
such as quadratic scoring rules based on the squared
errors of the predictions from the true observations
\citep{selten1998}, and scoring rules based on the
predictive cumulative distribution function
\citep{gneiting2007,gneiting2007calibration}.

While model selection takes a ``winner takes all approach'',
choosing the model that has the best score (e.g. the highest ELPD),
model averaging blends model predictions with weights
proportional to their predictive performance.
The canonical Bayesian paradigm is Bayesian model
averaging \citep[e.g.][]{hoeting1999}, which uses Bayes' rule
to estimate the posterior probablity of model $k$ given
the data, $p(M_{k} \mid y)$, and uses those posterior
model probabilities as weights to blend model predictions.
Bayesian model averaging only considers the in-sample performance,
however, and, in the limit of infinite data, will
assign the model with the best in-sample fit 100\%
weight \citep{yao2018,haines2024}.
Consequently, methods that estimate model weights
using measures of out-of-sample model performance
are generally favoured, reducing the risk of
generalization error to future data. Of the
various approaches, model stacking \citep{yao2018}
provides a principled method of estimating model
weights for Bayesian methods, which is highly
extensible \citep{yao2022,haines2024}. Stacking
finds model weights that maximize the average
expected log pointwise predictive densities, $\ELPD_i$, across
data points.

\section{A Bayesian workflow for casualty ILS}
\label{section:ledger-workflow}
With the basics of a Bayesian workflow now outlined, we now turn attention to the 
Bayesian workflow that we follow when securitizing casualty insurnace risk. 
To start, we emphasize that securitization of insurance programs involves two separate 
stages--the loss modeling stage and the deal modeling stage. The goal of loss modeling 
is to estimate ultimate losses for future accident years, including how long such 
ultimate losses take to develop. Outputs from loss modeling then serve as input to deal 
models, which are used estimate underwriting and investment cashflows given idiosyncratic 
deal terms that are negotiated between investors and insurers. Due to the idiosyncratic 
nature of deal modeling, our focus here is primarily on the loss modeling stage. However, due
to how important deal modeling is for securitization, we include a brief overview at the end 
to describe how to get from loss forecasts to the underwriting and investment cashflows that 
make up an ILS deal.

\begin{figure}[H]
    \centering
    \includegraphics[scale=.55]{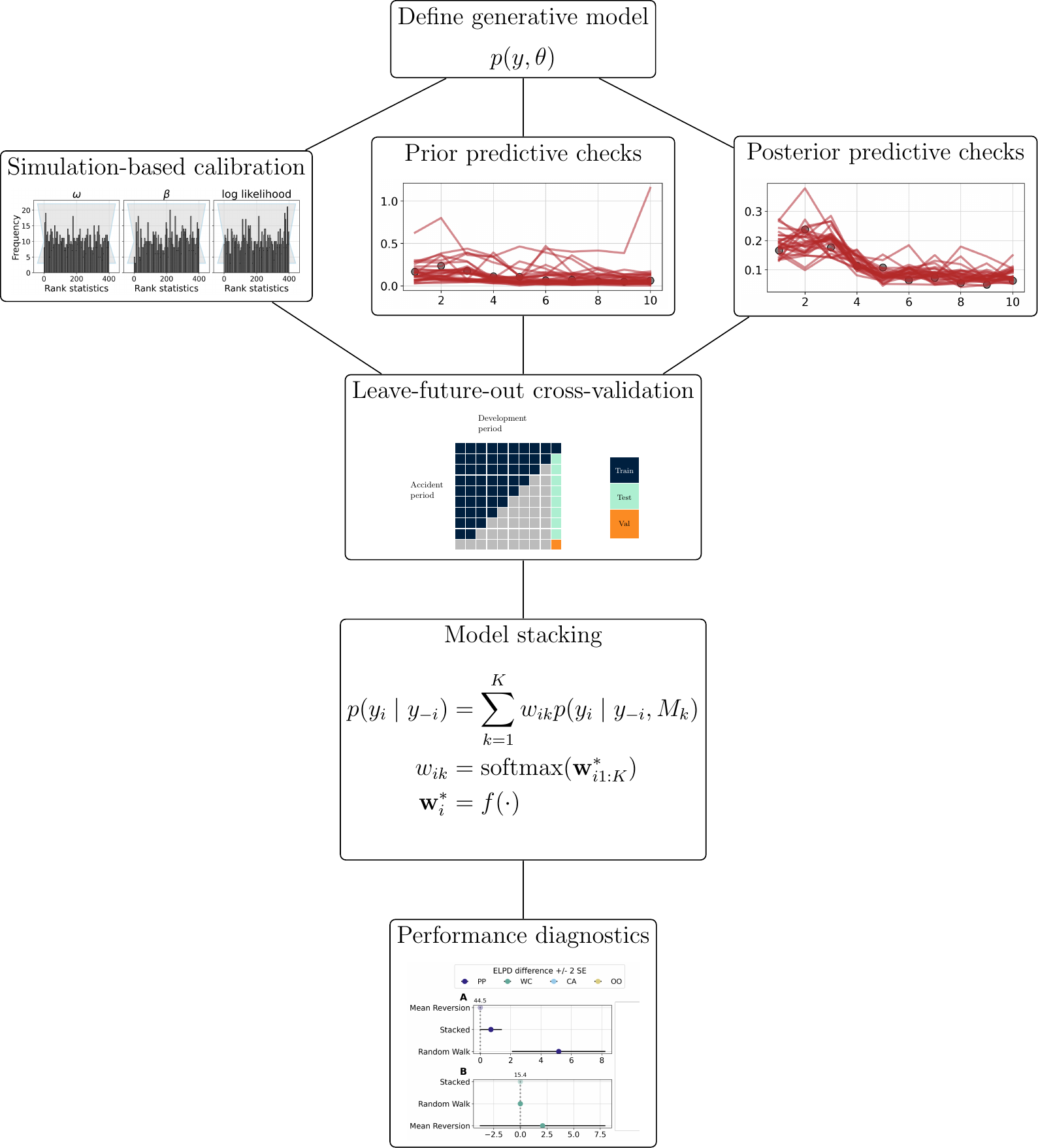}
    \caption{
        A Bayesian workflow for modeling casualty insurance loss development and 
        forecasting future losses -- the first step for securitization. In-depth 
        explanations for each step are described throughout Section
        \ref{section:ledger-workflow}. Here, $\theta$ represents all parameters in the 
        model, and $p(y, \theta)$ then represents the joint distribution of data and model
        parameters. Graphs in the workflow steps here are only included for illustration; we 
        dive into each workflow step in more detail in the following sections.
    }
	\label{fig:workflow}
\end{figure}

Figure \ref{fig:workflow} provides a visual representation of our Bayesian workflow. 
In the sections that follow, we walk through each step of the workflow in detail. Below, 
we start by defining generative models for loss development and forecasting. 

\subsection{Historical loss development} 
\label{section:loss-development}

There is an extensive body of literature on loss development in the actuarial 
literature, ranging from link-ratio models that directly capture multiplicative 
changes in cumulative losses between development lags 
\citep{mack1994, englandverrall2002} to parametric growth or decay models that capture 
cumulative losses directly \citep{zhang2012}. Link-ratio models are useful 
for capturing volatile patterns of change often observed across early development lags 
(which we refer to as the ``body'' of the loss triangle), but they are less identifiable 
for later development lags (commonly referred to as the ``tail'' of the loss triangle) 
due to the lack of observed data. Poor identifiability in the tail is problematic given 
our goal to extrapolate out to the ultimate loss for each accident year. Conversely, 
parametric models are better-suited for extrapolation, but they tend to be less flexible 
in the body. 

To resolve issues with using link-ratio versus parametric models in isolation, we 
follow \cite{tailfactors2013}, using variants of the chain-ladder and generalized 
Bondy models for the body and tail processes, respectively \citep{mack1993, tailfactors2013}.

Following the notation of \cite{goold2024}, we use
$\mathcal{Y}$ to denote the loss development triangle for an aggregated pool of insurance policies, defined by:

\begin{equation}
	\mathcal{Y} = \{y_{ij} : i = 1, ..., N; j = 1, ..., N - i + 1\}
    \label{eq:triangle}
\end{equation}

where $i = 1, ..., N$ indicates the accident year and $j = 1, ..., M$ indicates
the development lags. In real-world data, losses for a given accident year $i$ are
only known up to development lag $j = N - i + 1$, creating the triangular data 
structure that loss triangles are named for. Loss development models seek to predict 
the complement of $\mathcal{Y}$, or $\tilde{\mathcal{Y}}$, where the goal is to 
estimate $\tilde{y}_{i\infty}~\forall i = 1, ..., N$.

\subsubsection{Body development}

As described in Section \ref{section:chain-ladder-intro}, the chain-ladder model is 
a useful starting point for defining a loss development model more generally. 
However, the base model presented earlier can be extended to better capture how
uncertainty in ultimate losses changes as a function of development lag. Further, 
it needs modification to capture the body versus tail distinction we make above. Our 
fully-specified Bayesian variant of the chain-ladder model is defined as follows: 

\begin{align}
    \begin{split}
        y_{ij} &\sim \mathrm{Lognormal(\mu_{ij}, \sigma_{ij})}\\
        \mu_{ij} &= \log(\alpha_{j - 1} y_{ij-1})\\
        \sigma_{ij}^2 &= \exp(\gamma_{1} + \gamma_{2} j + \ln(y_{ij-1})),  \quad{\forall j \in (1, \tau]}\\
        \log \bm{\alpha}_{1:M - 1} &\sim \mathrm{Normal}(0, 1)\\
        \gamma_1 &\sim \mathrm{Normal}(-3, .25)\\
        \gamma_2 &\sim \mathrm{Normal}(-1, .1)\\
    \end{split}
    \label{eq:chain-ladder}
\end{align}

where $\bm{\alpha}$ is the vector of ``development factors'' that capture how losses
change across development lags. The prior on $\bm{\alpha}$ implies a median age-to-age factor
of 1.0 across development lags, with a lower 5\%ile of $\approx 0.2$ and upper 95\%ile of 
$\approx 5$. Therefore, the prior is quite diffuse, allowing for a wide variety of development
patterns that we might expect to see in real data. We purposefully choose diffuse priors over 
flat priors on age-to-age factors because the latter can often produce estimates that are completely
unrealistic as determined by prior predictive checks (see Section 
\ref{section:prior-posterior-predictions}).

In most ARIMA or state-space style time-series models, we expect that uncertainty would increase across 
time -- for loss development models, uncertainty propagation is a bit more complex. 
Intuitively, the further into the future we are when looking back at losses for a given accident year, 
the more claims are closed, and as a result we are more certain in the ultimate loss. To capture this 
behavior, we assume that variance in the ultimate loss follows a linear model with an intercept and 
slope captured by $\bm{\gamma} = (\gamma_1, \gamma_2)'$. We expect $\gamma_2 < 0$, indicating that 
losses vary around $\mu_{ij}$ less as development progresses. This decrease in variability is critical 
to capture the asymptotic behavior of losses across development -- without this heterogeneity mechanism, 
the model predictions will be too certain for early development lags and too uncertain for later 
development lags.

However, it is important to distinguish between observation noise (or $\sigma_{ij}^2$, 
the variance around $\mu_{ij}$) and parameter uncertainty (uncertainty in 
$\bm{\alpha}$ and $\bm{\gamma}$). Observation noise arises from epistemic uncertainty in loss 
development, whereas parameter uncertainty stems from our limited knowledge of the true values of 
$\bm{\alpha}$ and $\bm{\gamma}$. Because each $\mu_{ij}$ is a function of $\alpha_{j-1}$ and the 
previous observation $y_{ij-1}$ (or $\hat{y}_{ij-1}$ if the previous loss is itself a prediction), 
uncertainty compounds across development. This compounding effect is particularly clear when using the 
model for ``forecasting'' beyond observed development years.

Due to the two-stage nature of our loss development workflow, the chain-ladder model is 
only fitted to losses where $j \le \tau$, where $\tau \in {2,...,M}$ is an integer chosen by an 
analyst based on their knowledge of how long the body development process should be active. Generally, $\tau$ 
should be lower if the program is quick to develop (reaching the tail process for lower development 
lags), and higher if the program is slow to develop.

Finally, it is worth noting that the lognormal specification above could be replaced by a number of 
other distributions, including gamma, inverse-gamma, ex-Gaussian, and more. If frequency data is 
available, a frequency-severity model could be used to parse out different components of the 
underlying losses. We use the lognormal distribution here for simplicity, as our main goal is to 
illustrate how models from various stages of the workflow fit together.

\subsubsection{Tail development}
\label{section:tail}

Next, for the ``tail'' model, we follow from the actuarial literature \citep{tailfactors2013}, 
using a model based on the Bondy method. Our Bayesian ``generalized Bondy'' model is specified as 
follows: 

\begin{align}
    \begin{split}
        y_{ij} &\sim \mathrm{Lognormal(\mu_{ij}, \sigma_{ij})}\\
        \mu_{ij} &= \log(\alpha_{ij} y_{ij - 1})\\
        \alpha_{ij} &= \omega^{\beta^{j}}\\
        \sigma_{ij}^2 &= \exp(\lambda_{1} + \lambda_{2} j + \ln(y_{ij-1})), \quad{\forall j \in [\rho_1, \rho_2]}\\
        \log \omega &\sim \mathrm{Normal}^{+}(0, 1)\\
        \log \frac{\beta}{1 - \beta} &\sim \mathrm{Normal}(-2, .5)\\
        \lambda_1 &\sim \mathrm{Normal}(-3, .25)\\
        \lambda_2 &\sim \mathrm{Normal}(-1, .1)\\
    \end{split}
    \label{eq:bondy}
\end{align}

where the model form is identical to the chain-ladder model defined above, except the 
link ratios $\bm{\alpha}$ are now determined by a parametric decay model as 
opposed to being unconstrained free parameters. In the decay model, the asymptote 
$\omega$ is constrained to have a lower bound of 1.0 due to the truncated normal prior, which 
captures our assumption that losses will themselves asymptote to some 
ultimate value as develoment progresses. $\beta$ then controls the speed of decay, which 
we expect to be higher for more ``long-tailed'' lines of business (e.g., worker's 
compensation). However, values for $\beta$ too close to 1.0 can produce degenerate 
behavior where losses only asymptote at extreme development lags. Therefore, we center 
the prior such that decay in the resulting link ratios matches the timescale we expect 
(i.e., years- to decades-long decay). Unlike the chain-ladder model, the generalized 
Bondy model is fitted to only the window of development lags $j \in [\rho_1, \rho_2]$, 
where $(\rho_1, \rho_2) \in {2,...,M}$, $\rho_1 < \rho_2$, are chosen by an analyst based 
on where the tail process is assumed to begin and end. 

After fitting both the chain-ladder and generalized Bondy models, predictions are made 
by forward simulation starting from the left edge of the loss triangle for each 
accident year (i.e. $y_{i,1}$). The chain-ladder model is used to generate 
posterior predictions up to only $j = \tau$, and the generalized Bondy model is then 
used to generate posterior predictions from for all $j > \tau$ out to a development lag 
$j$ that is sufficiently large as to be practically indiscernable from $j = \infty$. 
These posterior predictive distributions for 
$\tilde{y}_{i,\infty} \forall i = 1, ..., N$ are then our ``developed ultimate
losses''.

\subsection{Future ultimate loss forecasting}
\label{section:forecasting}

Loss development models produce predictions for the ultimate losses associated with each 
historic accident year (or experience period more generally), yet the underlying risk 
being securitized is based on the future performance of a pool of policies yet to be 
written. Typically, securitization involves the ultimate losses in the next one to three accident 
years, or $\tilde{y}_{i,\infty} \forall i = (N+1, N+2, N+3)$. To solve this problem, we use 
the developed ultimate losses from the loss development model (i.e. 
$\tilde{y}_{i,\infty} \forall i = 1, ..., N$) as input to a time-series model used 
to forecast ultimate losses for future accident years. Bayesian models
are especially useful for this step, as they allow for us to easily incorporate 
uncertainty in the loss development model posterior predictions into the forecasting 
model.

The state-space modeling framework is particularly well-suited for loss forecasting, as
it allows for us to develop models that are informed by expert intuition 
regarding how losses evolve over time. In practice, we use a variety of different 
forecasting models dependent on use-case. For this example, we will overview a 
state-space random walk and state-space mean reversion model that illustrate our basic 
workflow. 

Before defining the models, we will introduce new notation. Although we find it 
useful for loss development to be directly on the losses, we find it easier to 
work with loss ratios for the forecasting step. Below, we use $r$ to indicate the loss 
ratio, such that 
$r_{ij} = \frac{y_{ij}}{p_i}~\forall i = 1, ..., N, \forall j = 1, ..., M$, where $p_i$ 
is the premium volume for accident year $i$.  

\subsubsection{A basic random walk model}

The base random walk model is defined as the following state-space model \citep{hyndman2018}: 

\begin{align}
    \begin{split}
        r_{i\infty} &\sim \mathrm{Lognormal}(\eta_{i}, \sigma_{i})\\
        \eta_{i} &\sim 
        \begin{cases} 
            \mathrm{Normal}(\eta_0, \epsilon), & \text{if } i = 1\\ 
            \mathrm{Normal}(\eta_{i-1}, \epsilon), & \text{otherwise}\\ 
        \end{cases} \\
        \sigma_{i}^2 &= \exp(\gamma_{1})^2 + \exp(\gamma_{2})^2 / \sqrt{p_i}\\
        \log \epsilon &\sim \mathrm{Normal}(-0.5, 1)\\
        \eta_0 & \sim \mathrm{Normal}(0, 1)\\
        \bm{\gamma}_{1:2} &\sim \mathrm{Normal}(-2, 1)
    \end{split}
    \label{eq:random-walk}
\end{align}

where $\bm{\eta} = (\eta_1, \dots, \eta_N)'$ is a latent (in log-scale) ultimate loss ratio that 
drifts across accident years with deviations between years proportional to $\epsilon$. The ``observed'' 
ultimate loss ratio ($r_{i\infty}$) then follows a lognormal distribution with observation 
noise $\sigma_{i}$ that is a function of earned premium for the given accident year ($p_i$) such that 
as earned premium increases, observation noise decreases. 

\subsubsection{Adding mean reversion}

The base state-space model is appealing because it allows us to separate observation 
noise from the ``true'' latent variability in the underlying ultimate loss ratios. The 
distinction is important, particularly in cases where the premium volume of a program
changes significantly over time, or when extending the model to capture effects that
influence the latent (but not observation) process. One example that comes up 
frequently in our own work is the addition of a mean-reverting mechanism to the random 
walk process:

\begin{align}
    \begin{split}
        \eta_{i} &\sim 
        \begin{cases} 
            \mathrm{Normal}(\mu(1-\phi) + \eta_0\phi, \epsilon), & \text{if } i = 1\\ 
            \mathrm{Normal}(\mu(1-\phi) + \eta_{i-1}\phi, \epsilon), & \text{otherwise}\\ 
        \end{cases} \\
        \mu &\sim \mathrm{Normal}(-1, 1)\\
        \log \frac{\phi}{1 - \phi} &\sim \mathrm{Normal}(0, 1)\\
    \end{split}
    \label{eq:mean-reversion}
\end{align}

where $\mu$ is the latent mean that the random walk process reverts back to, and $\phi$
indicates the strength and direction of the reversion process. All other model terms 
are identical to the random walk model defined above. Note that the mean reversion 
process is only on the latent log loss ratio scale, and the observation equation is 
unchanged. The motivation for mean reversion on the latent loss ratios is 
straightforward -- the underlying insurance program is being managed by a team that is 
constantly updating their expectations regarding how loss ratios will develop for the 
associated policies. The management team can then influence how future loss ratios unfold 
by changing underwriting practices, increasing or decreasing premium rates, and more. The 
end result is that ultimate loss ratios tend toward some target value as management makes 
decisions to ensure that the program is successful. 

\subsubsection{Adding measurement error}
\label{section:measure-error}

The forecasting models above are defined with $r_{i\infty}$ as the target. However, in 
real-world scenarios, we do not know $r_{i\infty}$ -- instead, we have the posterior 
predictions from the loss development models, or $\tilde{y}_{i\infty}$ (transformed to the 
loss ratio scale per $\tilde{r}_{i\infty} = \frac{\tilde{y}_{i\infty}}{p_i}~\forall i = 1, ..., N$). 
To account for uncertainty in these posterior predictions in the context of the forecasting 
model, we can add an error-in-variables measurement error model \citep{stefanski2000}
to the forecasting models that takes the posterior means 
$\mathbb{E} \left[ \tilde{r}_{i\infty} \right]~\forall i = 1, ..., N$ and posterior 
standard deviations 
$\mathbb{SD} \left[ \tilde{r}_{i\infty} \right]~\forall i = 1, ..., N$ as input and
models the ``true'' underlying ultimate loss ratio ${r'}_{i\infty}$. For both the 
random walk and mean reversion models, this involves the following modification to the 
likelihood expression:

\begin{align}
    \begin{split}
        {r'}_{i\infty} &\sim \mathrm{Lognormal}(\eta_{i}, \sigma_{i})\\
        \mathbb{E} \left[ \tilde{r}_{i\infty} \right] &\sim 
        \mathrm{Lognormal}(\mu_\xi, \sigma_\xi)\\
        \mu_\xi &= \log \frac{{r'}_{i\infty}^2}{\sqrt{{r'}_{i\infty}^2 + \mathbb{SD} \left[ \tilde{r}_{i\infty} \right]^2}}\\
        \sigma_\xi &= \sqrt{\log \left( 1 + \frac{\mathbb{SD} \left[ \tilde{r}_{i\infty} \right]^2}{ \mathbb{E} \left[ \tilde{r}_{i\infty} \right]^2}\right)}\\
    \end{split}
    \label{eq:measure-error}
\end{align}

Here, $\mu_\xi$ and $\sigma_\xi$ are the mean and standard deviation terms capturing
the relationship between the true (log) ultimate loss ratios ${r'}_{i\infty}$ and the 
observed posterior means from the development model posterior predictions. The true 
ultimate loss ratios are then used in the likelihood expression as the target for 
forecasting. 

Note that ${r'}_{i\infty}$ in equation \ref{eq:measure-error} is now a model parameter 
with its own prior. Above, the lack of prior implies a uniform distribution, which 
implies that the true ultimate loss ratio could take on any value. Of course, we often 
have quite a bit of information on what the distribution of ultimate loss ratios should 
be. Given the empirical mean ($\mathbb{E} \left[ r \right]$) and standard deviation 
($\mathbb{SD} \left[ r \right]$) of ultimate loss ratios that we have observed in 
historic data, we can specify a prior on ${r'}_{i\infty}$ that helps constrain the model 
to realistic values (see Section \ref{section:backtest} for an example):

\begin{align}
    \begin{split}
        {r'}_{i\infty} &\sim \mathrm{Lognormal}(\mu_{{r'}_{i\infty}}, \sigma_{{r'}_{i\infty}})\\
        \mu_{{r'}_{i\infty}} &= \log \frac{\mathbb{E} \left[ r \right]^2}{\sqrt{\mathbb{E} \left[ r \right]^2 + \mathbb{SD} \left[ r \right]^2}}\\
        \sigma_{{r'}_{i\infty}} &= \sqrt{\log \left( 1 + \frac{\mathbb{SD} \left[ r \right]^2}{\mathbb{E} \left[ r \right]^2}\right)}\\
    \end{split}
    \label{eq:measure-error-prior}
\end{align}

where $\mu_{{r'}_{i\infty}}$ and $\sigma_{{r'}_{i\infty}}$ are akin to $\mu_\xi$ and 
$\sigma_\xi$ in equation \ref{eq:measure-error}, representing the mean and standard 
deviation of the historic (log) ultimate loss ratios used to derive the informed prior. 
We find that such data-informed priors can often help with model diagnostics, including
both convergence of the sampler and out-of-sample model performance metrics such as RMSE 
and ELPD.

In practice, there are many potential measurement error assumptions one can make, and 
the above is just one example. In the current context, lognormal measurement error is 
reasonable given that the development model predictions are generated from a lognormal
distribution. An alternative approach would be to jointly model the loss development 
and forecasting stages such that there is no need for an intermediate summarization step
and subsequent measurement error modeling (akin to \cite{goold2024}, who jointly models
the body and tail stages of loss development). However, in practice we find that it is 
useful to separate the loss development from forecasting stages, which allows for more
intuitive setting of priors on the individual models and easier investigation of model
behavior. Further, having separate stages keeps the flow of information forward such 
that the forecasting model does not influence development model parameter estimation. 

\subsection{Deriving data-driven priors}
\label{section:data-driven-priors}

When defining models in previous sections, we only made conceptual arguments for why
particular priors were chosen. Priors can have quite a big impact on ultimate loss forecasts, 
so it is important to choose them in a data-driven way. This is particularly true if the triangle 
has minimal history, as is often the case for new programs undergoing ILS vetting. For example, say 
we are assessing a commercial auto program for a securitization opportunity, and the program 
has only existed for the past 5 years. With only 5 years of development, we may not 
have enough data to fit the tail portion of the loss development model (see 
Section \ref{section:tail}). Even if we were able to fit an adequate tail model, we 
would be left with only 5 accident years worth of ultimate loss ratios to feed into the 
rather complex forecasting models in Section \ref{section:forecasting}. In such cases, 
relying on models with uninformative or weakly informative priors can produce ultimate 
loss ratio predictions with uncertainty intervals far too wide to make ILS viable. 
Conversely, if we set priors that are too informative to combat low-data scenarios, we 
run the risk of biasing our model parameters, resulting in inaccurate and overconfident 
predictions that make an ILS deal appear more (or less) appealing than it is in 
reality. 

Our solution to this problem is to use hierarchical Bayesian analysis \citep{gelman2006hier}
to derive a set of data-driven priors for each model and line of business that we commonly work with. 
Hierarchical Bayesian analysis involves specifying a hierarchical model where the model 
parameters for each individual loss triangle themselves follow a group-level 
distribution. The intuition behind hierarchical models is simple -- in the absence of 
any historical loss information for a given program, our best guess for the program's
development pattern would be the development pattern for similar programs. Hierarchical 
models allow for us to estimate such group-level distributions from a pool of programs, which 
we can then use as priors for new programs undergoing ILS vetting.

The rationale for hierarchical models extends beyond intuition. Hierarchical models are 
closely related to credibility theory, which has informed actuarial practice for over a century 
\citep{michelbacher1918}. In particular, the mathematical form of hierarchical pooling is identical 
to a common form of classical credibility weighting (c.f. equations 2 and 3 in \cite{gelman2006} to 
equations 25 and 24 in \cite{whitney1918}, respectively). In both cases, the relative proportion of 
within- versus across-program variance is used to pool a program-level estimate toward the 
group-level average such that programs with higher volatility are more strongly pooled toward the 
group average and \textit{vice versa}.

As an example, the hierarchical variant of the random walk model (excluding measurement error) is 
as follows:

\begin{align}
    \begin{split}
        r_{gi\infty} &\sim \mathrm{Lognormal}(\eta_{gi}, \sigma_{gi})\\
        \eta_{gi} &\sim 
        \begin{cases} 
            \mathrm{Normal}(\eta_{g0}, \epsilon_g), & \text{if } i = 1\\ 
            \mathrm{Normal}(\eta_{gi-1}, \epsilon_g), & \text{otherwise}\\ 
        \end{cases} \\
        \sigma_{gi}^2 &= \exp(\gamma_{g1})^2 + \exp(\gamma_{g2})^2 / \sqrt{p_{gi}}\\
        \log \epsilon_g & \sim \mathrm{Normal(\epsilon_{\mu}, \epsilon_{\sigma})}\\
        \eta_{g0} & \sim \mathrm{Normal}(\eta_{\mu0}, \eta_{\sigma0})\\
        \bm{\gamma}_{g(1:2)} &\sim \mathrm{Normal}(-2, 1)\\
        \epsilon_{\mu} &\sim \mathrm{Normal}(-2, 0.5)\\
        \log \epsilon_{\sigma} &\sim \mathrm{Normal}(-2, 0.5)\\
        \eta_{\mu0} &\sim \mathrm{Normal}(-1, 0.5)\\
        \log \eta_{\sigma0} &\sim \mathrm{Normal}(-2, 0.5)
    \end{split}
    \label{eq:hierarchical}
\end{align}

where $g = 1,2,...,G$ is the program index within a group of similar programs. The model is 
functionally equivalent at the individual loss triangle level, but now 
instead of having hard-coded priors on the triangle-level parameters, the group-level parameters 
(or hyper-parameters) that are estimated from the data. For example, $\log \epsilon_g$ now follows 
a group-level normal distribution with log mean and log standard deviations $\epsilon_{\mu}$ and 
$\epsilon_{\sigma}$, respectively. This scheme does require specifying priors on the group-level 
parameters, but their influence is rather minimal if the number of programs $G$ is sufficiently 
high enough to allow for precise estimation of the group-level parameters. We are often in this 
situation, where we have many example triangles for each line of business we work with, 
yet each triangle has minimal observations individually. The hierarchical model allows 
for partial pooling of information across triangles such that information on parameters
for each triangle informs the group-level parameter, in turn informing all other 
triangle-level parameters. 

In practice, it is not computationally efficient to fit a hierarchical model to all 
historic triangles in addition to the triangle of interest when we are analyzing data
for a new program. Therefore, we often hierarchical models in a two-step fashion. First, 
we define the grouping variable of interest, which determines which triangles will be 
included in a given hierarchical model. Typically, we group triangles by line of 
business (e.g., whether a triangle is from a commerical auto, private passenger, 
general liability, or worker's compensation insurance program). Next, we limit our 
focus to only recent triangles (e.g., if it is 2024, we may only consider triangles 
from 2014 onward). Once the group is defined, we fit the hierarchical model to the data
within the group (and undergo model validation exercises described below in Section 
\ref{section:validation}) and then summarize the group-level parameters. For example, 
once we have posteriors from the hierarchical model for $\epsilon_{\mu}$ and 
$\epsilon_{\sigma}$, the prior on the non-hierarchical model (equation 
\ref{eq:random-walk}) becomes $\log \epsilon \sim \mathrm{Normal}(\mathbb{E}\left[\epsilon_{\mu}\right], \mathbb{E}\left[\epsilon_{\sigma}\right])$. When analyzing a new program that fits into one of the 
pre-defined lines of business, we use the corresponding group-level parameter summaries as priors 
in the non-hierarchical variant of the model, thus obtaining the hierarchical pooling (or credibility
weighting) described above in a dynamic, data-driven way. Note that this approach assumes that we can 
estimate the group-level parameters with sufficient precision, which is often the case when we have many 
programs within a given line of business. In cases where we have fewer programs within a group, it is 
sensible to scale the empirical group-level standard deviation (i.e. $\mathbb{E}\left[\epsilon_{\sigma}\right]$) by some factor based on the number of programs in the group to ensure that the prior is not 
overly informative.

\subsection{Model validation with simulated data}
\label{section:validation}

With the loss models defined, the next step is to validate the models using the
suite of techniques from the Bayesian toolbox as described throughout Section
\ref{section:bayes-workflow}.

\subsubsection{Simulation-based calibration}

Typically, we perform SBC for each stage of our modeling pipeline separately, for each 
new model we develop. For illustration, here we apply SBC to just the loss development 
modeling stage. 

We simulated data from the loss development body and tail models using the exact priors
specified when introducing the models above. We simulated 1000 separate datasets, each 
generated from a randomly sampled set of parameters from the afforementioned prior 
distributions. We set the body development lag cutoff to $\tau = 5$, and the tail 
development lag window to $\rho = (6, 10)$. Each dataset was generated to have 10 
accident years and 10 development lags, thus matching the dimensions of a typical 
real-world loss triangle (the first accident year has 10 development lags of observed 
losses, second accident year has 9, third has 8, etc., until the 10th accient year has 
just 1). 

For each of the 1000 simulated parameter sets and associated loss triangles, we fit the 
loss development model with $\tau = 5$ and $\rho = (6, 10)$, using the same set of 
priors used in the simulation. For each fit, we drew 1000 posterior samples for each of 
4 indpendent MCMC chains. We then generated predictions out to the 10th development 
lag, thus filling out the lower diagonal of the triangle with posterior predictions. 
Next, we thinned the posterior samples such that we only kept overy 10th sample, 
resulting in a total of 400 posterior samples with negligible autocorrelation. 

\begin{figure}[H]
    \centering
    \includegraphics[scale=.5]{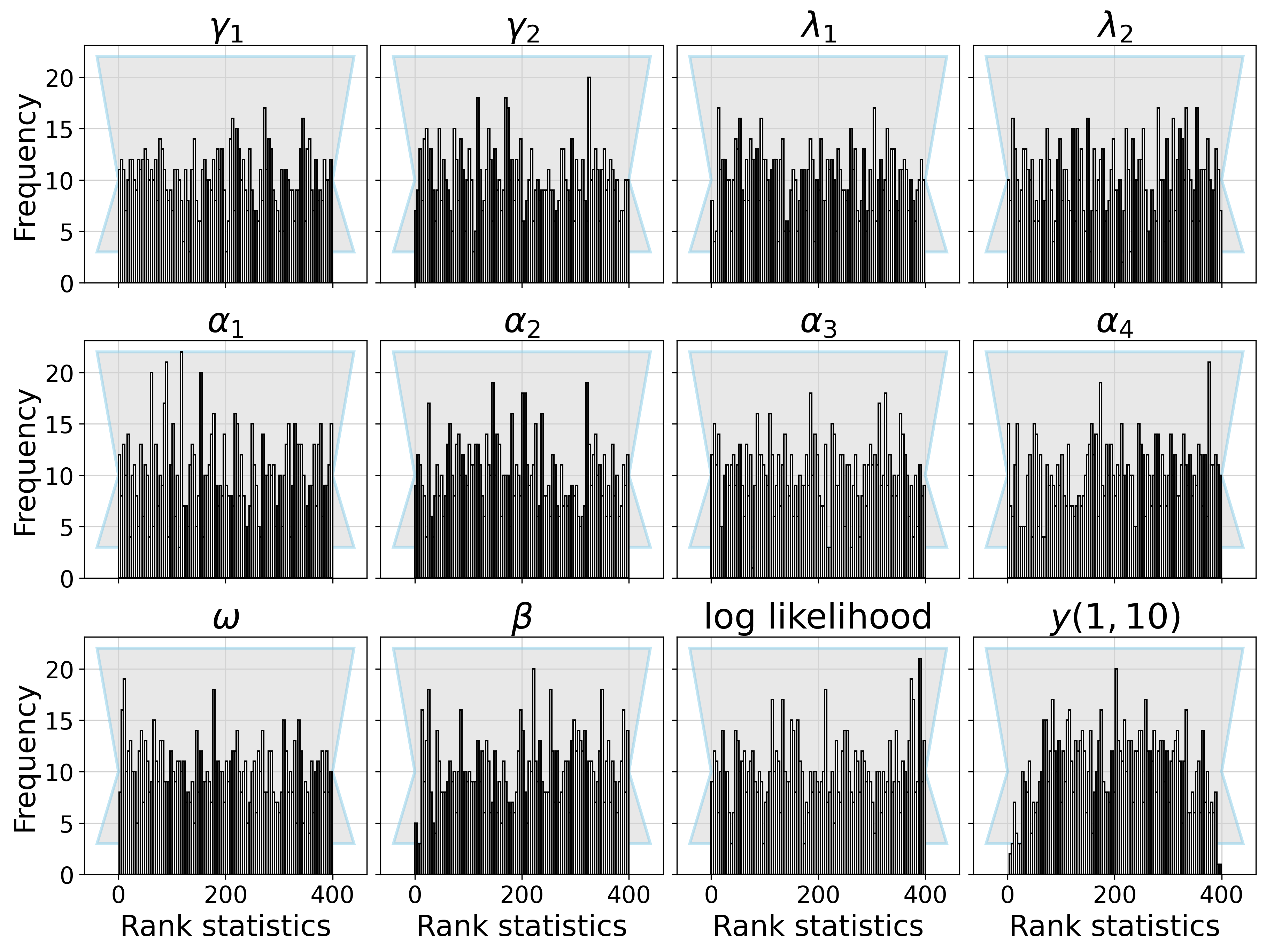}
    \caption{
        Simulation-based calibration rank histograms. For each model, we sampled
        4000 draws from the posterior distribution, and thinned the samples by 10 to 
        remove any autocorrelation, meaning a maxmimum rank statistic of 400.
        Uniformly distributed ranks indicate good calibration. Error ribbons indicate 99\% 
        uncertainty intervals for the target uniform distribution. Histogram bars falling 
        outside of the intervals are then evidence of mis-calibration.
    }
	\label{fig:ranks}
\end{figure}

Finally, for each dataset and model fit pair, we computed the rank of the true 
parameter values and losses in the associated posterior distributions. Figure \ref
{fig:ranks} shows the distribution of these ranks across all 1000 dataset and model fit 
pairs. Rank distributions for all parameters are indistinguishable from uniform 
distributions, indicating proper specification and good parameter recovery. A slight 
exception is for the posterior predictions, which show an inverted U shape. The inverted 
U shape signifies too much uncertainty in the posterior predictions. However, the ranks 
are almost entirely within the bounds we would expect for data generated from a uniform 
distribution, so the extra uncertainty is minimal. 

\subsubsection{Prior and posterior predictive checks}
\label{section:prior-posterior-predictions}

\begin{figure}[H]
    \centering
    \includegraphics[scale=.5]{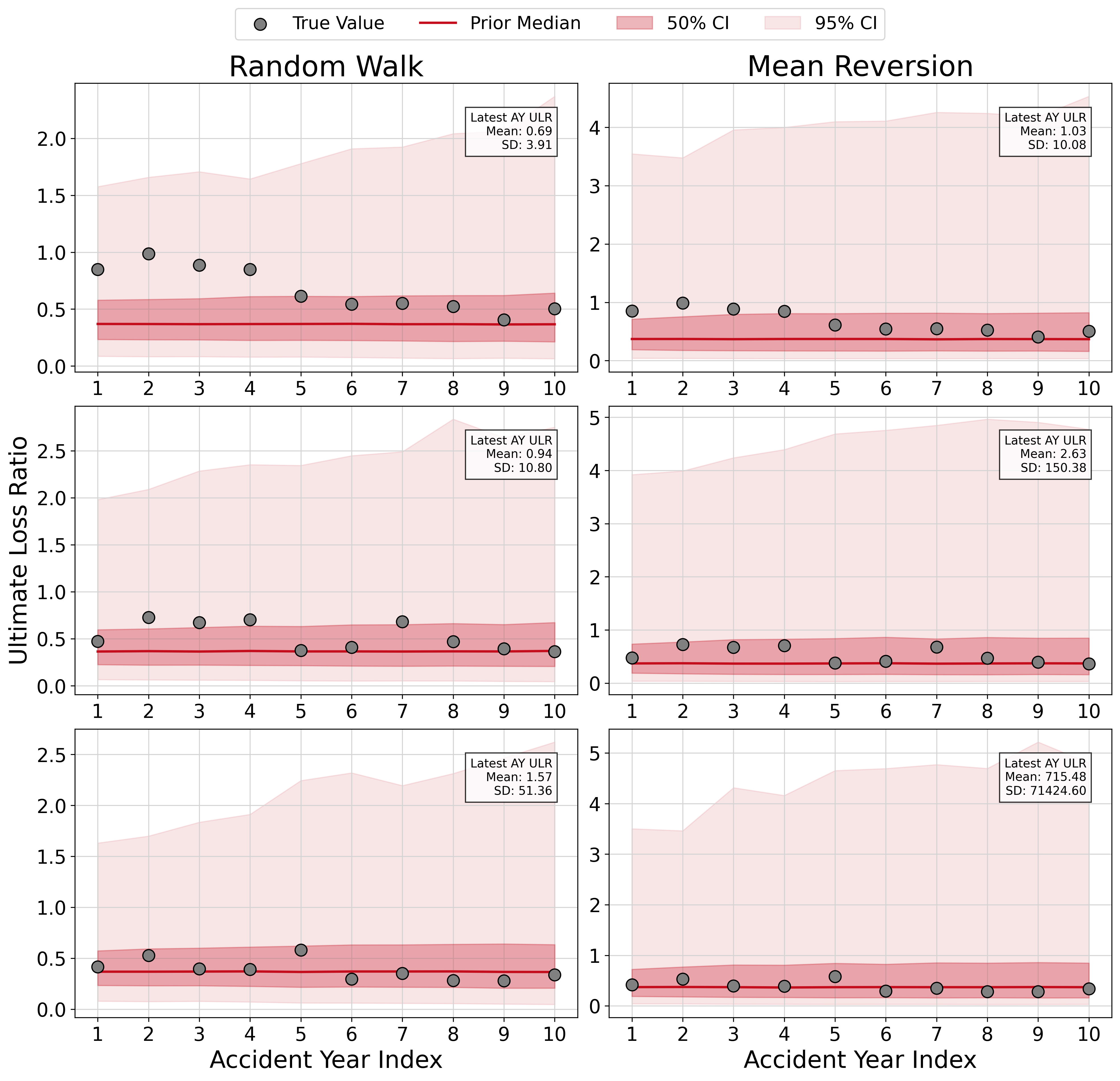}
    \caption{
        Prior predictions for three random commercial auto programs from the 
        \cite{meyers2015} dataset. Shaded intervals indicate 50\% and 95\% credible intervals, and 
        points are true losses at $y_{i(10)}$. 
    }
	\label{fig:prior}
\end{figure}

Figure \ref{fig:prior} illustrates prior predictive checks on three different example 
commercial auto programs from \cite{meyers2015}. Note that we focus only on the 
forecasting model here for brevity, using the same priors defined in Section 
\ref{section:forecasting}. To do so, we fit the loss development model to the triangle, 
then used the posterior predictions on the ultimates as input to the forecasting model 
as described in Section \ref{section:measure-error}. The prior predictive simulations 
in Figure \ref{fig:prior} are generally well-behaved, generating a wide range of loss
patterns, most of which do not show the afforementioned ``exploding'' or``crashing'' 
behaviors alluded to in Section \ref{section:chain-ladder-intro}. Compared to the real 
losses, the simulations tend to show much more variability, but this is expected before 
the model is conditioned on the real losses.

\begin{figure}[H]
    \centering
    \includegraphics[scale=.5]{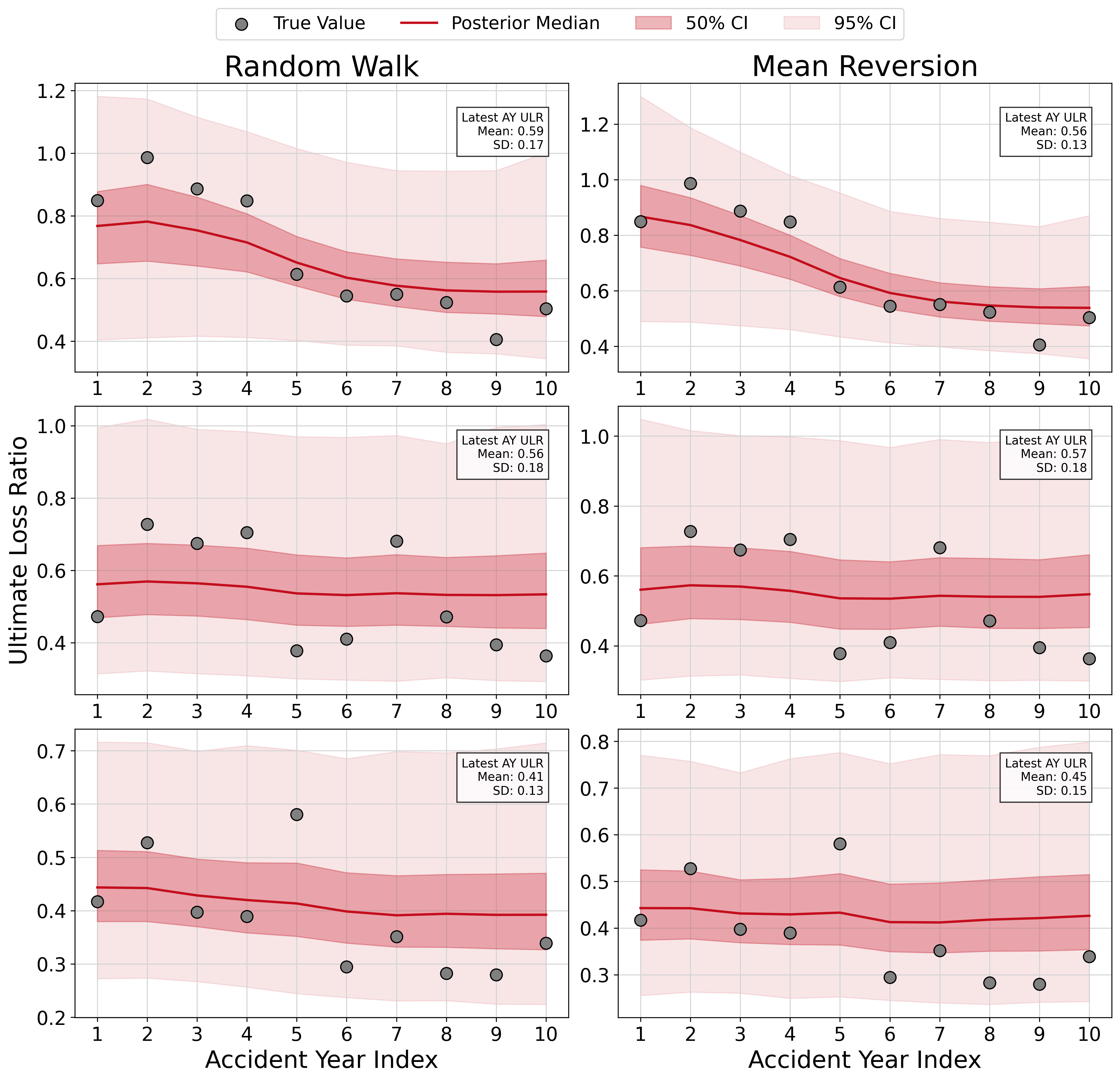}
    \caption{
        Posterior predictions for three random commercial auto programs from the 
        \cite{meyers2015} dataset. The programs here are the same as in Figure 
        \ref{fig:prior}. Shaded intervals indicate 50\% and 95\% credible intervals, 
        and points are true losses at $y_{i(10)}$. 
    }
	\label{fig:posterior}
\end{figure}

Figure \ref{fig:posterior} then shows posterior predictive checks for the same data in 
Figure \ref{fig:prior}, but after fitting the model to the real losses. The predictions 
now follow the real losses with much more precision, capturing the slow-moving drift in 
losses across accident years. Both the random walk and mean reversion models 
show a slight amount of over-estimation of losses for later accident years. However, 
some mis-esimtation of the true losses is expected given that the forecasting models 
are trained only on the development model predictions as input. 

\subsection{Backtesting workflow using empirical data}
\label{section:backtest}

A model can pass all model validation checks but then fail to perform well on 
real-world prediction tasks given the wide range of different loss triangles that we 
encounter from day-to-day. Therefore, large-scale out-of-sample tests are crucial to 
determine if a given model (or model pipeline) is suited for production use. Further, 
we often have multiple competing models, where each model makes different assumptions
regarding how losses change over time. For example, equation \ref{eq:mean-reversion} 
assumes mean reversion toward some latent ultimate loss, whereas equation 
\ref{eq:random-walk} assumes no such latent ultimate loss or reversion process. An 
added benefit of conducting out-of-sample tests is that we can take advantage of modern
Bayesian stacking methods that combine predictions from competing models, allowing for 
a combined performance better than the best fitting individual model. In the 
following sections, we overview our out-of-sample backtesting workflow and our use of
stacking to improve model performance.  

Often referred to as ``backtesting'', we use a form of exact leave-future-out cross 
validation (see \cite{vehtari2012} for a review of different cross-validation schemes).
Backtesting in general is highly dependent on the specific data that one has available 
for use when testing. For our internal backtests, we license access to public statutory 
filings data via a vendor. This dataset contains information from insurance programs across 
over a dozen different lines of business, altogether comprising over 10,000 triangles, most with 
history dating back to the 1980's. 

\begin{figure}[H]
    \centering
    \includegraphics[scale=1]{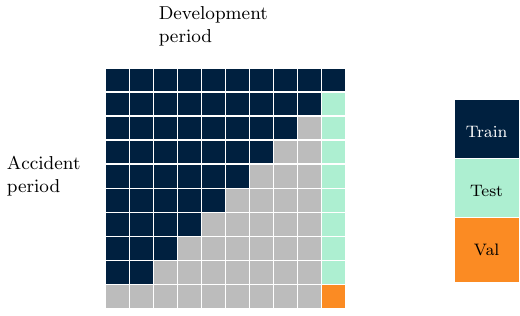}
    \caption{
        A schematic of the backtest procedure. Train datapoints are used to fit the 
        loss development model, which is used to make ultimate loss predictions out to 
        the right edge of the triangle. Predicted ultimate losses are then used as input 
        for the forecasting models, which is used to generate out-of-sample predictions 
        on both the Test and Validation datasets. 
    }
	\label{fig:backtest}
\end{figure}

For demonstration purposes, here we will rely on an open dataset described by 
\cite{meyers2015}, which contains about 50 loss triangles for each of four different 
lines of business, including private passenger auto, worker's compensation, commercial 
auto, and general liability lines. Each triangle in the Meyers dataset is 10 accident 
years by 10 development lags, which allows for us to fit our loss development to loss 
forecasting pipeline on each triangle exactly once, using the loss from the most recent 
accident year, development lag pair as the future, left-out datapoint for testing 
purposes (see Figure \ref{fig:backtest}). Note that this scheme assumes 
$y_{i\infty} = y_{i(10)}~\forall i = 1, ..., N$, which may not be true in practice for
longer-tailed lines of business. However, given data constraints for this demonstration,
we believe this is a reasonable assumption. We use $y_{i\infty}$ and $y_{i(10)}$ 
interchangeably in sections that follow.

For the loss development stage, we fit the chain-ladder and generalized Bondy models to 
each Meyers triangle indpendently using the specifications outlined in equations 
\ref{eq:chain-ladder} and \ref{eq:bondy}, respectively. For the shorter-tailed private 
passenger and commerical auto lines, we set the chain-ladder training threshold to 
$\tau = 4$, and the generalized Bondy training window to $\rho = (5, 10)$. For the 
remaining longer-tailed lines, we set $\tau = 6$ and $\rho = (4,10)$. For all lines, we 
then generated posterior predictions out to $\tilde{y}_{gi(10)} \forall i = 1, ..., 9$ for each 
triangle $g$ following the procedure in Section \ref{section:tail}, which were used as 
ultimate losses for input to the forecasting models. 

For the forecasting stage, we then fit both the random walk (equation 
\ref{eq:random-walk}) and mean reversion (equation \ref{eq:mean-reversion}) models 
hierarchically (see equation \ref{eq:hierarchical}) across all triangles within each of 
the four lines of business in the Meyers dataset. Both models also used the measurement 
error specification in equation \ref{eq:measure-error}. We also used the measurement 
error prior defined in equation \ref{eq:measure-error-prior}. Because $r_{g1(10)}$ 
is the only ``observed'' ultimate loss ratio for each triangle given our backtest scheme, 
we took the mean and standard deviation of these observed loss ratios across triangles 
within each line of business, and then used these in place of 
$\mathbb{E} \left[ r \right]$ and $\mathbb{SD} \left[ r \right]$, respectively, in 
equation \ref{eq:measure-error-prior}. After fitting the models, we generated posterior 
predictions for each triangle for $\tilde{r}_{gi(10)} \forall i = 1, ..., 10$. 

Finally, to determine how sensitive our results are to the prior specifications, we ran the 
full backtesting workflow pipeline with two alternative prior specifications -- one with more diffuse 
priors, and one with more informative priors. We relegate details of this analysis to the appendix,
but include some results below in the main text for completeness.

\subsubsection{Stacking of predictive distributions}
\label{section:stacking}

Once backtesting is complete, we can use the out-of-sample performance metrics to 
further improve predictive performance. In practice, we often use hierarchical Bayesian 
stacking to do so, a form of stacking model that allows you to model weights conditional 
on pointwise covariates \citep{yao2022}. Having stacking weights conditional on 
covariates is particularly useful in cases where we believe that models will perform 
better in different contexts. Common examples include the program's line of business or 
earned premium for the current accident year -- different lines of business might 
be better explained by one model over another, and similarly for different accident years. 
Instead of simply choosing the best individual model for each context, stacking in such cases 
allows us to dynamically weight models when most appropriate.

For demonstration purposes, we will use the stacking model described by 
\cite{yao2018}, which estimates a single weight for each model. The model can 
be easily fit using \texttt{bayesblend} -- an open-source \texttt{Python} library that we 
have developed that implements various different stacking models \citep{haines2024}. To 
fit the stacking model, we first computed the posterior log likelihood values and 
posterior predictions for each datapoint in the test set as illustrated in Figure 
\ref{fig:backtest}. We did so for both the random walk and mean reversion forecasting 
models. Note that each posterior log likelihood in the test set is derived based on the 
true loss, which the models are not trained on. We then used the posterior log 
likelihoods and associated posterior predictions for each model as input to the 
\texttt{MleStacking} class in \texttt{bayesblend}, which outputs a blend of the 
posterior log likelihoods and predictions given the estimated model weights. 

\subsubsection{Model performance metrics}
\label{section:assess}

To assess model performance, we use a variety of out-of-sample metrics including 
(1) expected log pointwise predictive density (ELPD), (2) root mean squared error
(RMSE), and (3) predictive versus true percentiles. The use of multiple metrics helps 
to hedge against the shortcomings of any particular metric, giving us a fuller
understanding of how a model may perform on new programs. Note that all metrics are 
derived using the validation set as illustrated in Figure \ref{fig:backtest}. Focusing 
on the validation set is important because we are most interested in future accident 
year performance. Additionally, it is necessary for a fair comparison between the 
individual models and the stacked model given that the latter is trained on the test 
set performance. Below, we define each metric before presenting the results.

As described in Section \ref{section:model-compare-intro}, ELPD is estimated using 
predictions on out-of-sample predictions. In our case, out-of-sample predictions for each 
triangle include those where $i = 2, ..., N$. However, because we are primarily 
interested in future accident year performance, we focus our metric definitions only on 
the ``validation'' datapoint, where $i = N = 10$ (see Figure \ref{fig:backtest}). We 
define the validation set LPD for model as the log likelihood values for each validation 
datapoint, marginalized across the posterior samples $S$: 

\begin{align}
	\label{eq:lpd}
	\begin{split}
	\mathrm{LPD}^{\mathrm{val}}_{g}
        &= \log p(\tilde{y}_{g10(10)} \mid \mathcal{Y})\\		
        &= \log \int p(\tilde{y}_{g10(10)} \mid \theta)
			p(\theta \mid \mathcal{Y}) d \theta \quad \forall g = 1, ..., G\\
		&\approx \frac{1}{S} \sum_{s=1}^{S} 
			\log p(\tilde{y}_{g10(10)}^{(s)} \mid \mathcal{Y})
	\end{split}
\end{align}

The expected LPD (ELPD) is then defined as the sum of LPD values across the validation
datapoints, or across the different triangles $g$ in our case:

\begin{equation}
\label{eq:elpd}
	\mathrm{ELPD}^{\mathrm{val}} = \sum_{g=1}^{G} \mathrm{LPD}^{\mathrm{val}}_{g}
\end{equation}

Higher values of ELPD indicate better model performance. When comparing models, we
took the difference in pointwise LPD values between models and then computed the sum of 
the differences and their associated standard errors as described by 
\cite{vehtari2017}. This pointwise comparison is akin to a paired $t$-test on the LPD 
values across triangles. 
 
We defined pointwise out-of-sample RMSE as follows: 

\begin{equation}
	\label{eq:rmse}
	\mathrm{RMSE}^{\mathrm{val}}_{g} = \sqrt{\frac{1}{S} \sum_{s=1}^{S} (y_{g10(10)} - \tilde{y}_{g10(10)}^{(s)})^2} \quad \forall g = 1, ..., G
\end{equation}

Unlike ELPD, RMSE primarily penalizes inaccuracy, where predictions further from the 
true loss are increasingly penalized. Similar to ELPD, for model comparison we computed
the pairwise difference in RMSE across triangles, then computed the mean of these 
differences and their associated standard errors. 

Finally, we computed the percentile of the true value in the posterior predictive 
distribution for each out-of-sample validation datapoint. Formally, the pointwise 
percentiles are defined as: 

\begin{equation}
	\label{eq:percentiles}
	\mathrm{PERC}^{\mathrm{val}}_{g} = \frac{1}{S} \sum_{s=1}^{S} \mathbf{1}_{\{\tilde{y}_{g10(10)}^{(s)}, y_{g10(10)}\}}
\end{equation}

where $\mathbf{1}$ is an indicator function that returns 1 if the true value is lower
than the posterior sample and 0 otherwise: 

\begin{equation}
    \mathbf{1}_{\{X, Y\}} = \begin{cases} 
        1 & \text{if } X < Y \\
        0 & \text{else}
    \end{cases}
\end{equation}

The resulting percentiles should follow a uniform distribution if the model's posterior 
predictions are well-calibrated. By assessing the empirical distribution of predictive
percentiles, we can determine if the model is making loss predictions that are too low, 
too high, too certain, or too uncertain at the aggregate level. 

\subsubsection{Model performance results}
\label{section:model-compare}

\begin{figure}[H]
    \centering
    \includegraphics[scale=.5]{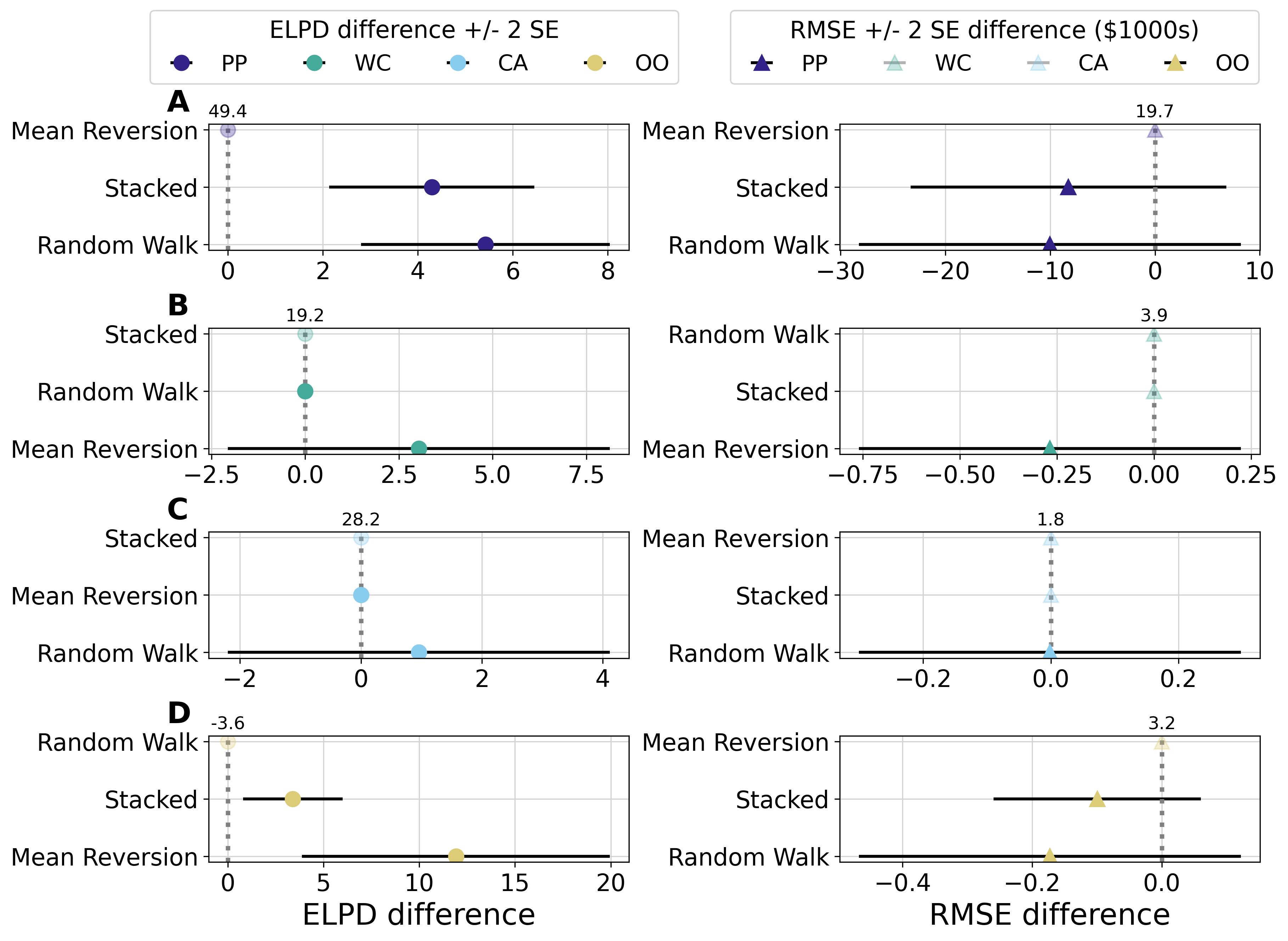}
    \caption{
        ELPD (left) and RMSE (right) differences (+/- 2 standard errors on the difference), 
        ordered by Validation dataset performance for each model and line of business. The 
        best-performing model is the reference model, shown at the top of each panel. 
        Absolute ELPD and RMSE values are displayed above. Positive ELPD differences with 
        an uncertainty interval that does not cross zero indicates a credible difference 
        at the 95\% level in favour of the top model. Negative RMSE differences with an uncertainty 
        interval that does not cross zero indicates a credible difference in favor of 
        the top model. Note that in cases where non-reference models do not have visible intervals, 
        the uncertainty intervals are just too narrow to be visible on the plot.
    }
	\label{fig:scores}
\end{figure}

Figure \ref{fig:scores} shows how the random walk, mean reversion, and stacked models 
compare in terms of both ELPD and RMSE within the four lines of business in the Meyers 
dataset. There is significant variability in which model performs best across different
lines of business, although the ELPD and RMSE metrics tend to agree with each other. 
Specifically, the mean reversion model tends to perform better for the private passenger 
auto (PP) and commercial auto (CA) programs, and the random walk model performs better 
for the general liability (OO) and worker's compensation (WC) programs. These findings 
align with our intuitive understanding of the different lines of business -- auto programs 
tend to develop very quickly, meaning that the feedback loop between a management team 
observing losses and then engaging in behaviors that change losses or premiums is rather 
short. Such dynamics are captured by the mean reversion mechanism in equation 
\ref{eq:mean-reversion}. Conversely, the losses underlying general liability and worker's 
compensation programs typically take many years to develop, making the aforementioned 
feedback loop longer and thus harder to act on. With enough historic data, it is possible 
to capture longer-term market cycle trends in such data, but the short term dynamics 
are better described by the random walk process in equation \ref{eq:random-walk}.

Across all lines of business, the stacked model is either similar to or better than the 
best individual candidate model within each line. The only exception is for general 
liability, where the ELPD metric slightly favors the random walk model over the stacked 
variant. Because the stacking model is trained on the test data and model performance 
is only estimated on the validation data (see Figure \ref{fig:backtest}), it is not 
surprising that the stacked model occasionally performs worse than individual candidate 
models. However, when looking across lines, the stacked model provides the best 
compromise in the sense that it is the model configuration that performs best in 
aggregate across contexts. 

\begin{figure}[H]
    \centering
    \includegraphics[scale=.5]{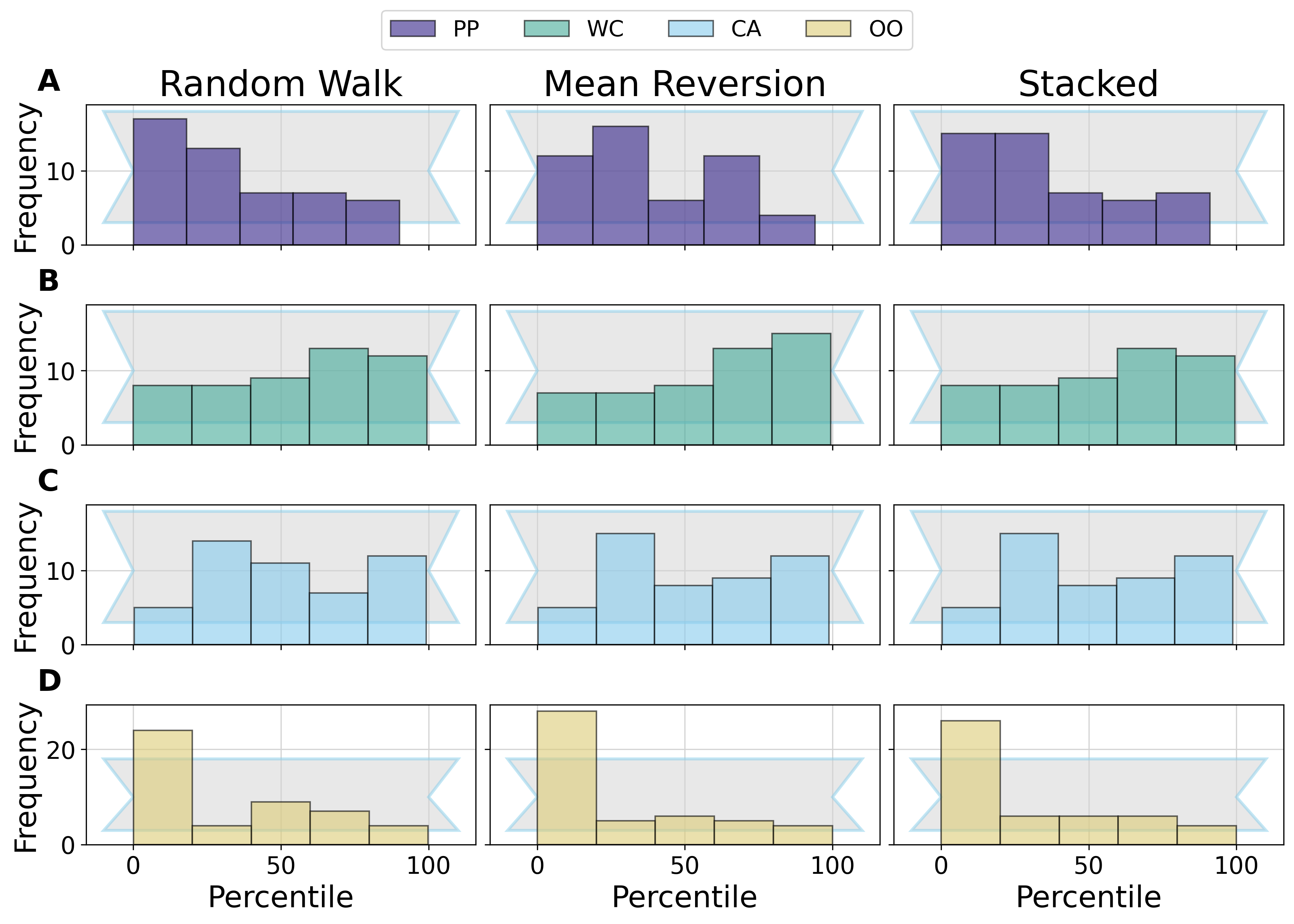}
    \caption{
        Percentiles of the true hold-out Validation dataset losses within the 
        corresponding posterior predictive distributions for each model and line of 
        business (panels A through D). Grey shaded regions provide the 99\% intervals 
        of a target uniform distribution, for reference. Right-skewed histograms indicate 
        under-estimation, left-skewed histograms indicate over-estimation, inverted-U 
        histograms indicate predictions that are uncertain, and U histograms indicate
        predictions that are too certain.
    }
	\label{fig:calibration}
\end{figure}

Figure \ref{fig:calibration} shows the model calibration results for each model within 
each line of business. Here, we see that calibration is good for commercial auto and 
worker's compensation lines, but calibration for private passenger auto and general 
liability lines reveals that loss predictions have an upward bias relative to true losses 
across all models. We tried various different modifications to the underlying models in 
an effort to improve calibration for these lines, including changes to the loss 
development training windows $\tau$ and $\rho$, different measurement error assumptions 
in the forecasting model (e.g., normal as opposed to lognormal error), changes to priors 
for both loss development and forecasting models, and different outcome distributions for 
the loss development and forecasting models (e.g., gamma as opposed to lognormal) -- all 
changes resulted in the same general pattern found in Figure \ref{fig:calibration}. 

Poor calibration should be taken quite seriously, as having accurate forecasts with 
proper uncertainty quantification is crucial for pricing in ILS deals. Therefore, the 
poor calibration for private passenger auto and general liability lines would indicate 
that the models presented are not suitable for production use in those lines, even if 
they perform well according to other tests. Of course, these particular results are 
highly dependent on the data used for backtesting purposes. The Meyers dataset is limited 
in that we only have a 10-year window of history for each underlying triangle, starting 
in 1988 and ending in 1997. This limited history means that all out-of-sample validation 
data is coming from accident year 1997 (see Figure \ref{fig:backtest}), which may not 
give a generalizable view of how models should be expected to perform for arbitrary 
accident years. In fact, exogenous factors such as the macroeconomic environment or general 
industry conditions -- neither of which are explicitly captured by models presented here -- 
could induce correlations in ultimate losses across programs within each line of business. When 
left unaccounted for, such correlations could contribute to poor calibration.

In practice, uncertainty regarding the generalizability of any given accident year is why 
we typically perform backtests using a rolling window approach, where the procedure in 
Figure \ref{fig:backtest} is iterated across overlapping 10-year accident year 
increments. However, this rolling window approach requires a dataset with much more data, 
and such rich datasets are not openly available in the insurance industry. Anecdotally, 
using our internal licensed data with much more history, we have found that calibration 
for the forecasting models presented here is generally good across lines of business, 
suggesting that the results in Figure \ref{fig:calibration} may in part be due to the 
idiosyncracies of accident year 1997. 

\subsection{From Forecasts to Cashflows}

Once we have selected a model configuration that can forecast ultimate losses for future accident years 
to a desired level of accuracy (including calibration, RMSE, etc.), the next step is to estimate 
cashflows that can be used to understand the profitability of the insurance program. There are two types 
of cashflows that are relevant for this step: (1) the ``underwriting'' cashflow, which captures 
primarily how losses are paid out across time for the future accident year being securitized; and (2) 
the ``investment'' cashflow, which captures interest on invested premium dollars (i.e. the ``float'') as 
well as bespoke deal structure terms (collateralization schemes, profit commission to cedants, etc.). 
For the current demonstration, we will focus primarily on the underwriting cashflow, as the investment 
cashflow is a deterministic function of underwriting cashflow (i.e. it is mostly an exercise in 
accounting) so is outside the scope the current work.

\subsubsection{Estimating future underwriting cashflows}
\label{section:estimate-cashflows}

There are many ways to estimate future underwriting cashflows, but for illustration we 
will take a simple approach here. Underwriting cashflows are comprised primarily of paid losses 
(outflow) versus premiums collected (inflow). Other inflows can include reinsurance or legal recoveries, 
and other outflows can include expenses related to claim settlement or management -- here we will only 
consider paid losses and premiums.

\begin{figure}[H]
    \centering
    \includegraphics[scale=.35]{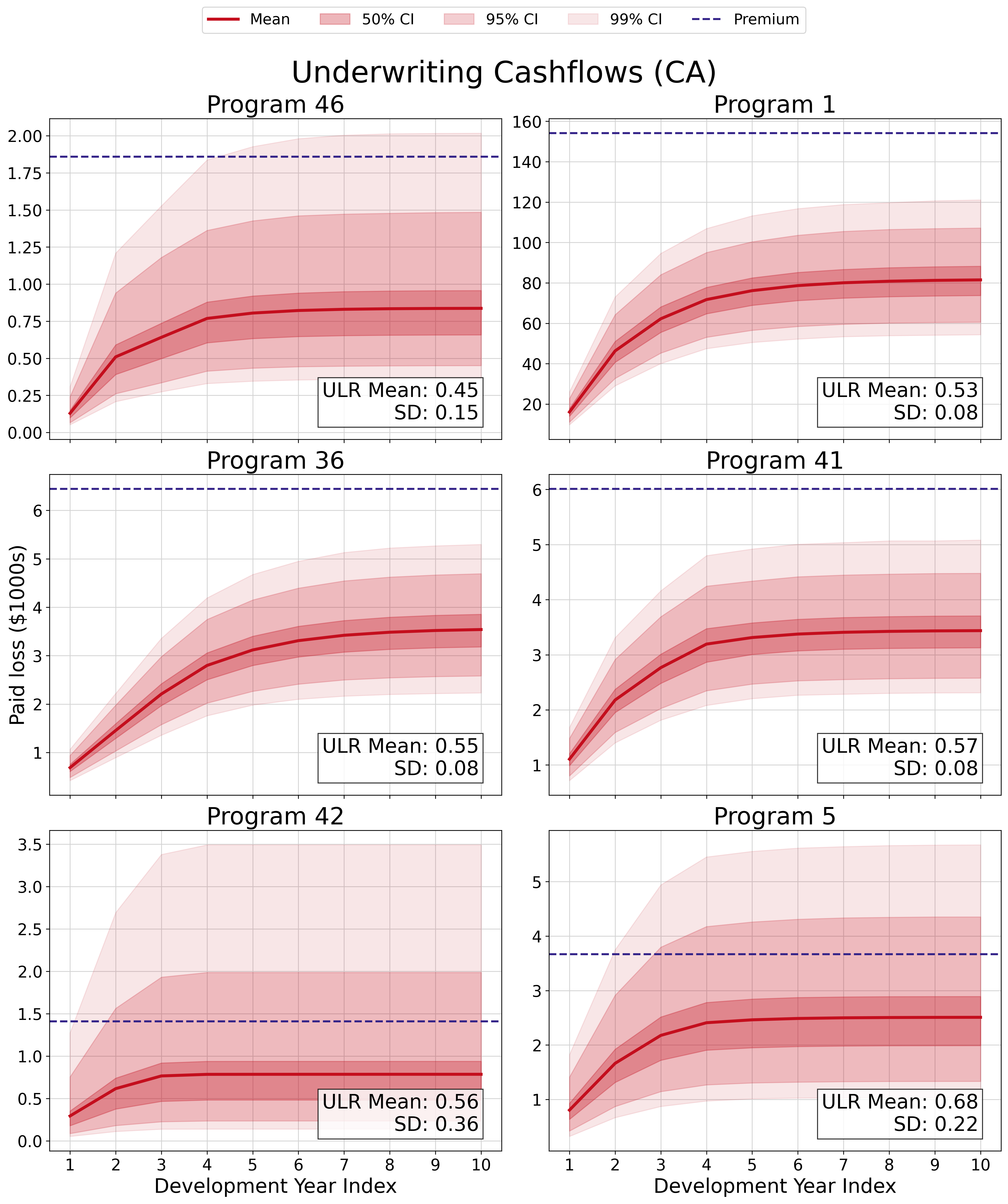}
    \caption{
        Underwriting cashflows for future accident years based on historic development 
        patters and forecasted ultimate losses for a random selection of commercial auto programs in the 
        Meyers data. Results are based on the development and forecasting models fitted during the 
        backtests described above. Underwriting cashflows were estimated per the walk-back method 
        described in equation \ref{eq:cashflows} using random programs selected from the Backtest 
        workflow (see Section \ref{section:backtest}) illustration. Forecasted ultimate losses were 
        based on the stacked predictions from the random walk (equation \ref{eq:random-walk}) and mean 
        reversion (equation \ref{eq:mean-reversion}) models, and age-to-age factors were obtained from 
        the chain-ladder (equation \ref{eq:chain-ladder}) and generalized Bondy (equation 
        \ref{eq:bondy}) models fitted to each program.
    }
	\label{fig:cashflow}
\end{figure}

Given the estimated age-to-age factors and the next year(s) ultimate loss 
estimate(s) from the development and forecasting models, respectively, we can walk back from the 
ultimate loss estimate to obtain estimates of future paid losses. Because we have full posterior 
distributions for the ultimates loss estimates and age-to-age factors, we can propagate both 
parameter and epistemic uncertainty from the development and forecasting models into the paid loss 
estimates by leveraging the MCMC samples. Assuming $\tilde{r}_{i\infty} = \tilde{r}_{i(10)}$ 
represents the ultimate loss ratio prediction for future accident year $i$ obtained from a 
forecasting or stacking model, which makes $\tilde{y}_{i\infty} = \tilde{y}_{i(10)}$ the corresponding 
ultimate loss estimate derived by multiplying the loss ratio by the premium associated with the 
same period, we can estimate future paid losses as follows:

\begin{align}
    \begin{split}
        \tilde{y}_{i(9)}^{(s)} &= \tilde{y}^{(s)}_{i(10)} / \alpha^{(s)}_{9}\\
        \tilde{y}_{i(8)}^{(s)} &= \tilde{y}^{(s)}_{i(9)} / \alpha^{(s)}_{8} \quad \forall s = 1, ..., S\\
        &\vdots\\
        \tilde{y}_{i(1)}^{(s)} &= \tilde{y}^{(s)}_{i(2)} / \alpha^{(s)}_{1},
    \end{split}
    \label{eq:cashflows}
\end{align}

where $\alpha^{(s)}_{j}$ is the $s$th posterior sample of the $j$th age-to-age factor. The result is a full posterior distribution of paid loss estimates for each 
future accident year $i$ (see Figure \ref{fig:cashflow}). Across development, these paid loss estimates, 
relative to premiums, correspond to the underwriting cashflow associated with the future accident year.

\subsubsection{Underwriting to investment cashflows}
\label{section:investment-cashflows}

Because underwriting cashflows only inform us as to how insurance policies are expected to 
be paid out over time, they are not sufficient for pricing ILS deals. An ILS fund can be thought of 
as a trust account comprised of several additional inflows and outflows that determine its overall 
performance. In addition to the inflows and outflows associated with underwriting cashflows, the primary 
investment cashflows include investment income and initial investor capital (inflows) as well as 
expenses associated with managing or brokering the security (outflows). Although we will not cover 
technical details on investment cashflows here, it is worth discussing the general structure of an ILS 
fund to understand the relationship between underwriting and investment cashflows.

Initially, the fund consists of the total premiums collected from policyholders plus the 
investor capital. The capital requirement is determined as some proportion of the expected claim 
payments, ensuring that the fund has sufficient reserves to cover potential losses. Notably, the 
investor capital is considered leveraged because interest is earned on the total amount of investor 
capital plus the premiums collected. Because the capital contributed by the investor is often 
significantly less than the amount of premium ceded by the insurer, it is common for investors to 
receive more than 2x leverage on investment returns. This leverage means that even if the interest is a 
low single-digit yield, such as from bonds or treasury bills in the current market, investors often 
realize a much higher internal rate of return (IRR) on capital. The longer it takes for 
claims to come in and be paid out (i.e. the longer it takes for losses to develop), the longer the 
investor's capital remains leveraged.

Of course, many of the above details are negotiable and depend on the specific deal 
structure. For example, the capital requirement may be higher or lower if the underlying insurance 
porfolio is concentrated or diversified, respectively. Similarly, expenses can vary widely depending on 
the complexity of the deal and the services provided by the ILS fund manager. For example, managers of 
the insurance program may be incentivised with bonuses based on the loss ratio of the program, affecting 
underwriting cashflows. Additionally, the ILS broker may receive a fee based on the total amount of 
capital raised, affecting investment cashflows.

\section{Working example}
\label{section:working-example}
\begin{figure}[H]
    \centering
    \includegraphics[width=0.85\textwidth]{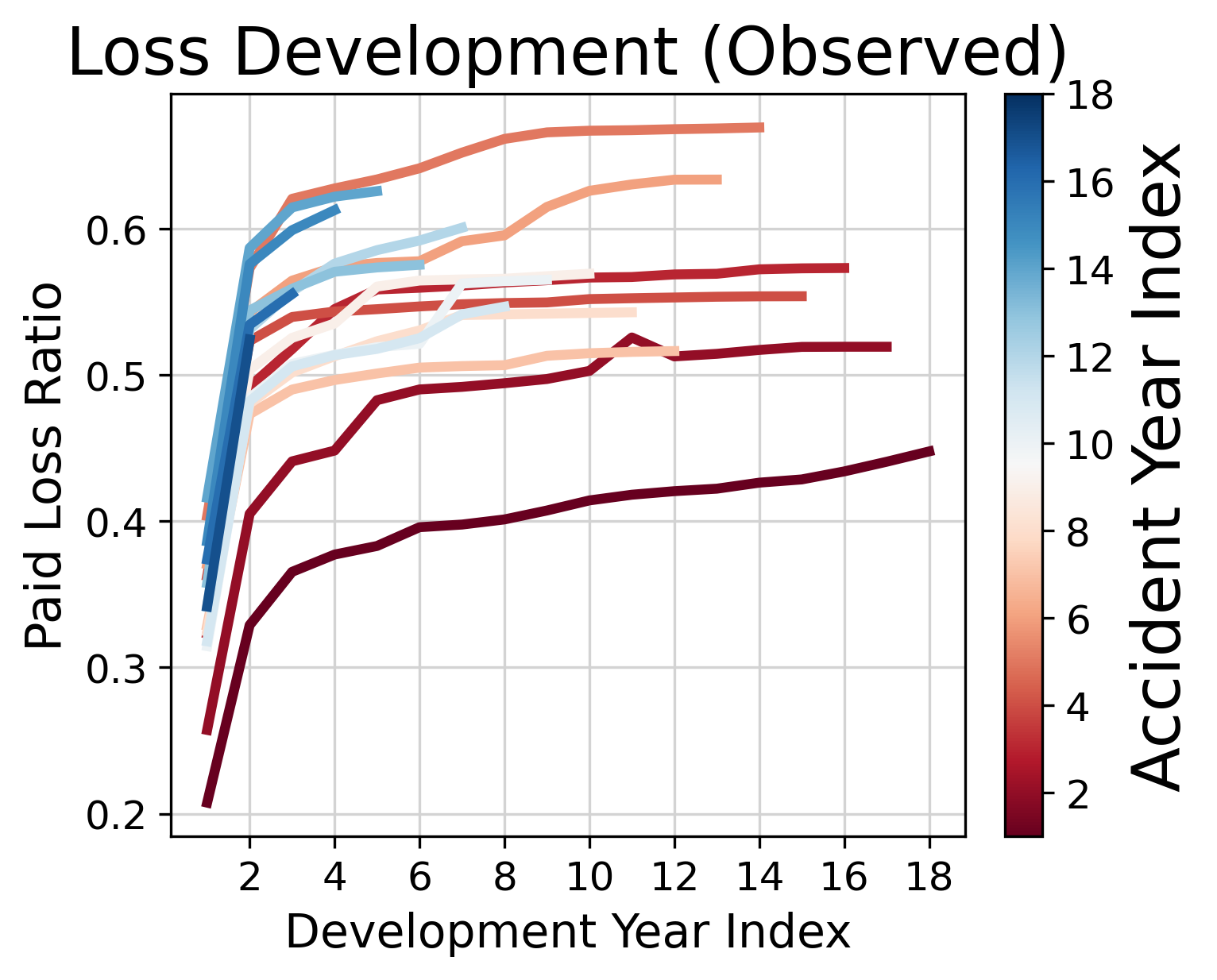}
    \caption{Observed loss ratio development patterns from the Swiss liability program in \cite{balona2020}, indicating a generally long-tailed structure with variability in early development years.}
    \label{fig:observed-oo-balona-2020}
\end{figure}

In this section, we apply the full Bayesian workflow presented in Section 
\ref{section:ledger-workflow} to the Swiss private liability loss development triangle from 
\cite{balona2020} (refer to Tables 23-24). These data are originally from \cite{gisler2009}, but Balona 
(2020) add simulated premium data to the triangle that is needed for our workflow, so we will refer to 
\cite{balona2020} as the source from hereon out. The Swiss private liability triangle has 18 years of 
history spanning from 1979 to 2015, which originate from a major Swiss insurer. Note that the exact 
figures have been adjusted for privacy purposes--we report values directly from \cite{balona2020} in 
thousands of dollars, but emphasize that the scaling is arbitrary. As seen in Figure 
\ref{fig:observed-oo-balona-2020}, the data exhibit the typical long-tailed, early-lag volatility that 
characterizes many insurance programs undergoing casualty securitization. As such, it makes for an 
insightful example of how to apply the Bayesian workflow in practice.

\subsection{Loss development}
\label{section:body-tail-modeling}

Following the historical loss development workflow from Section \ref{section:loss-development}, 
we separate loss development into a ``body'' process (via the Bayesian chain-ladder model in 
equation \ref{eq:chain-ladder}) and a ``tail'' process (via the generalized Bondy model in 
equation \ref{eq:bondy}). Because private liability falls under the more general umbrella of general 
liability insurance, we specify the the chain-ladder and generalized bondy training thresholds 
using similar values as for those used to backtest the general liability line above (see Section 
\ref{section:backtest}). Specifically, we set $\tau = 6$ and $\rho_1, \rho_2 = (4,15)$. We also used
the same prior distributions as specified in equations \ref{eq:chain-ladder} and \ref{eq:bondy} for the 
chain-ladder and generalized Bondy models, respectively.

Note that the upper threshold we set for the tail model ($\rho_2$) is higher here because the 
Swiss liability triangle has 18 years of history available, whereas the Meyers data used for 
backtests only had 10 years of history. Further, unlike for the backtest, we do not set the 
upper tail threshold to the maximum possible given the training data (i.e. $\rho_2 = 18$). 
Including the full training data means that tail development factors for the most recent 
development lags are informed only by the most historic accident years where data are available. 
By not fitting the tail model to the most recent development lags, we allow for the model to be 
informed only by development patterns where we have multiple accident years, which can make the 
tail estimates more robust to noise that may otherwise be present in 1 or 2 early accident 
years.

\begin{figure}[H]
    \centering
    \includegraphics[width=0.85\textwidth]{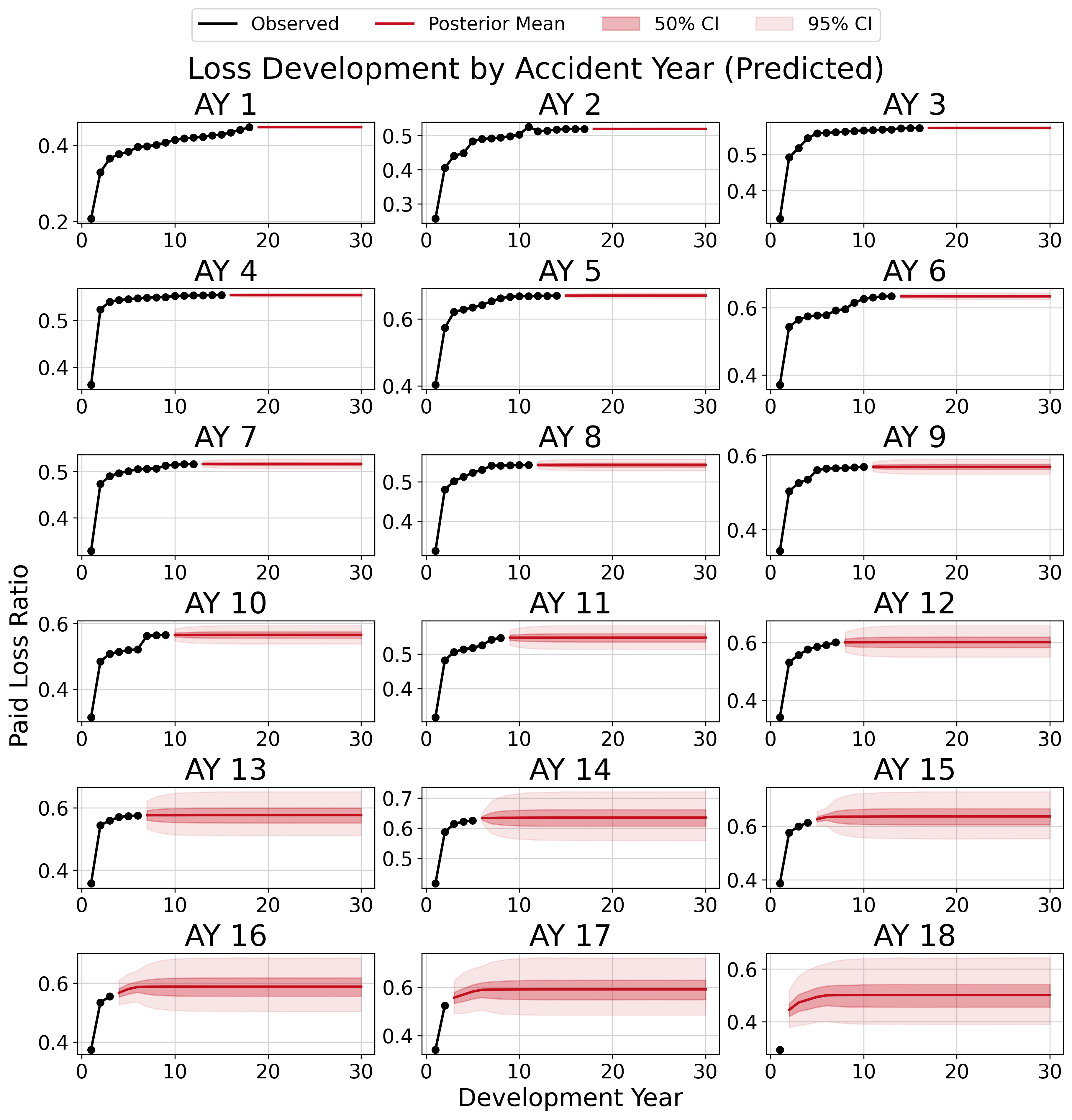}
    \caption{Observed loss ratio development patterns along with posterior predictions from loss development models for the Swiss liability program in \cite{balona2020}. The predicted ultimate loss ratio means and standard deviations (i.e. predicted means and standard deviations at development year 30) are used as input to the forecasting model described in Section \ref{section:ultimate-forecast}. The shaded regions correspond to the 50\% and 95\% credible intervals for the predictions.}
    \label{fig:post-pred-oo-balona-2020}
\end{figure}

Figure \ref{fig:post-pred-oo-balona-2020} shows posterior predictions from the development models on the 
Swiss private liability program, faceted by accident year to better visualize the uncertainty in the 
predictions. Generally, we see that the model predictions are mostly flat after 6-7 years
of development. It is also clear that the model is more uncertain for more recent accident years, 
which is expected given we only have one or more observed years of development for the latest accident 
years. Overall, the importance of the development model is most readily observed for later accident 
years, where the model captures the expected increase in cumulative losses across early development 
periods before converging toward the ultimate loss ratios for each year.

\subsection{Ultimate loss forecasting}
\label{section:ultimate-forecast}

Given the predicted ultimate losses obtained for historic accident years (see Figure 
\ref{fig:post-pred-oo-balona-2020}), the next step is to predict ultimates for future accident years. 
For the current example, we will use the same forecasting models as described in Section 
\ref{section:forecasting}. Specifically, we fit both the random walk (equation \ref{eq:random-walk}) and 
mean reversion (equation \ref{eq:mean-reversion}) models, both with the measurement error specification 
in equation \ref{eq:measure-error}. After generating one-year-ahead predictions from each fitted model, 
we then blend together the posterior predictive distributions to arrive at the final one-year-ahead 
ultimate loss ratio prediction.

For blending the forecasting model predictions, we used the same \texttt{MleStacking} model as in our 
backtests (see Section \ref{section:stacking}). Recall that in our backtests, we trained a separate 
stacking model for each of the four lines of business in the \cite{meyers2015} dataset. To blend the 
Swiss liability triangle predictions, we used the stacking model trained on the general liability subset 
of the Meyers data. The \texttt{MleStacking} model only assigns a single static weight to each 
forecasting model, which in our case produced weights of 0.55 for the random walk model and 0.45 for the 
mean reversion model. Given these weights, the blended ultimate loss ratio distribution is generated by 
sampling the posterior predictive distributions from each forecasting model in proportion to each
model's respective weight.

Finally, for our forecasting model priors, we used data-driven priors as described in Section 
\ref{section:data-driven-priors}. Specifically, when conducting the backtests for each line of business 
in the Meyers dataset (see Section \ref{section:backtest}), we saved the posterior distributions 
from the hierarchically estimated forecasting models (see equation \ref{eq:hierarchical}), summarized 
the group-level parameters for each line of business, and used these summaries as the priors for each 
forecasting model fitted to the Swiss liability triangle:

\begin{align*}
    \eta_{\text{init}} &\sim \mathcal{N}(-0.92, 0.41) \tag{Random Walk} \\
    \epsilon &\sim \mathcal{N}(-2.17, 0.22) \\
    \gamma_1 &\sim \mathcal{N}(-2.18, 0.96) \\
    \gamma_2 &\sim \mathcal{N}(-2.13, 0.92) \\
    ~\\
    \eta_{\text{init}} &\sim \mathcal{N}(-1.49, 0.86) \tag{Mean Reversion} \\
    \epsilon &\sim \mathcal{N}(-2.66, 0.15) \\
    \mu &\sim \mathcal{N}(-0.65, 0.52) \\
    \phi &\sim \mathcal{N}(1.27, 0.95) \\
    \gamma_1 &\sim \mathcal{N}(-2.23, 0.91) \\
    \gamma_2 &\sim \mathcal{N}(-2.24, 0.86) 
\end{align*}

\begin{figure}[H]
    \centering
    \includegraphics[width=0.85\textwidth]{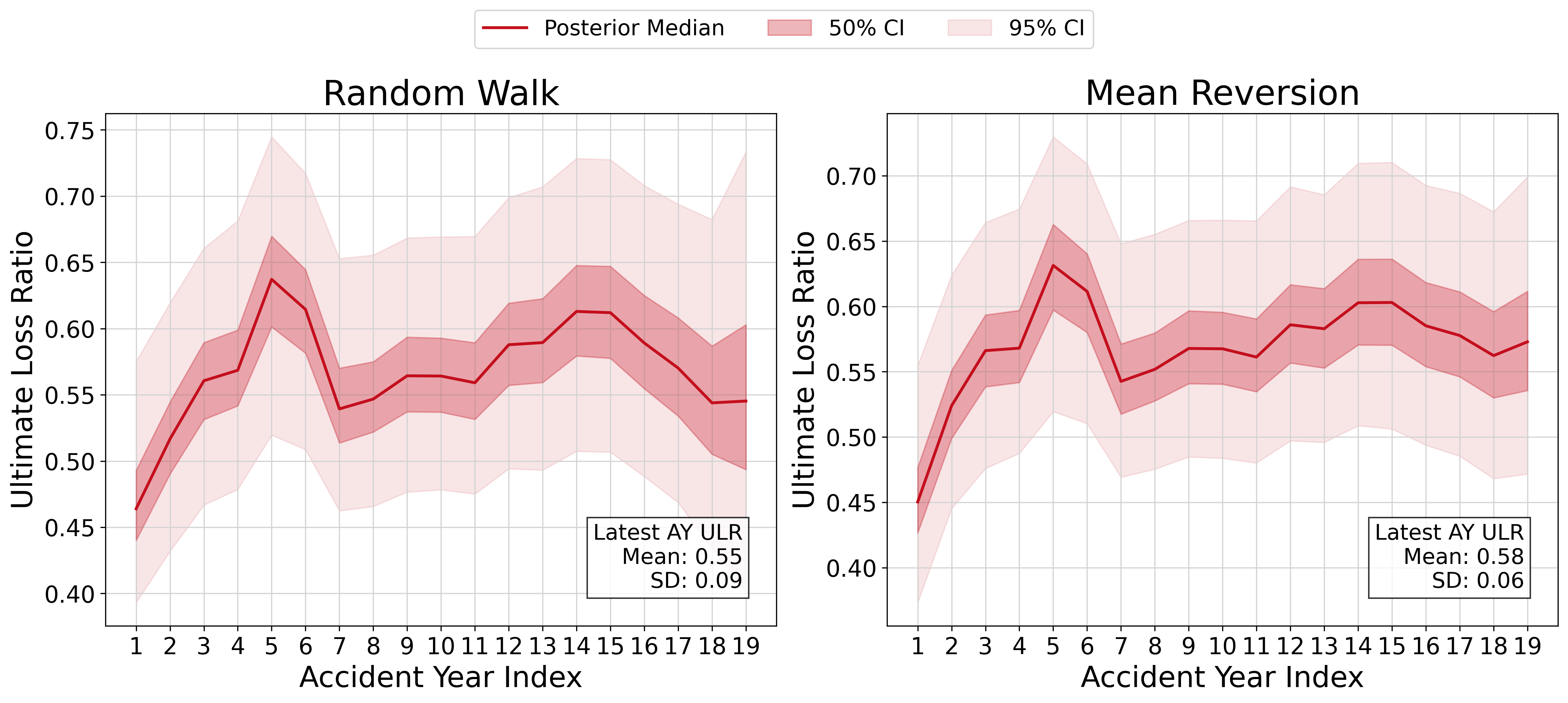}
    \caption{Posterior predictive distributions for each forecasting model across accident years for the Swiss liability triangle from \cite{balona2020}. The shaded regions correspond to the 50\% and 95\% credible intervals for the predictions.}
    \label{fig:posterior-predictions-OO-balona-2020}
\end{figure}

Figure \ref{fig:posterior-predictions-OO-balona-2020} shows the posterior predictive distributions for 
each forecasting model across accident years, including the one-year-ahead predictions. The 
models capture the same general patterns, although the mean reversion model tends to produce predictions 
with less uncertainty than the random walk model. This occurs because the mean reversion process pulls 
year-to-year predictions toward the mean of the historical data, which reduces overall variability in 
the predictions. Further, the mean reversion model produces a slightly higher ultimate loss ratio mean 
than the random walk model. The difference in both the expectation and uncertainty of predictions across 
models highlights the importance of model blending, allowing us to combine models as opposed to 
selecting just one. 

\subsection{Underwriting Cashflows}
\label{section:cashflow-modeling}

The final step we will cover in our working example is to predict one-year-ahead underwriting cashflows 
given the forecasted ultimate loss ratio predictions and the historic loss development patterns. Here, 
we use the walk-back method outlined in equation \ref{eq:cashflows}. As explained in Section 
\ref{section:estimate-cashflows}, the walk-back method starts from the one-year-ahead ultimate loss ratio
obtained from the forecasting models, then uses the age-to-age factors from the loss development model 
(Section \ref{section:body-tail-modeling} above) to iteratively predict back to the first development 
lag. This process results in a full time-series of posterior predicted cashflows for the one-year-ahead
accident year, which represent the paid losses expected to emerge for the future accident year across 
time.

\begin{figure}[H]
    \centering
    \includegraphics[width=0.85\textwidth]{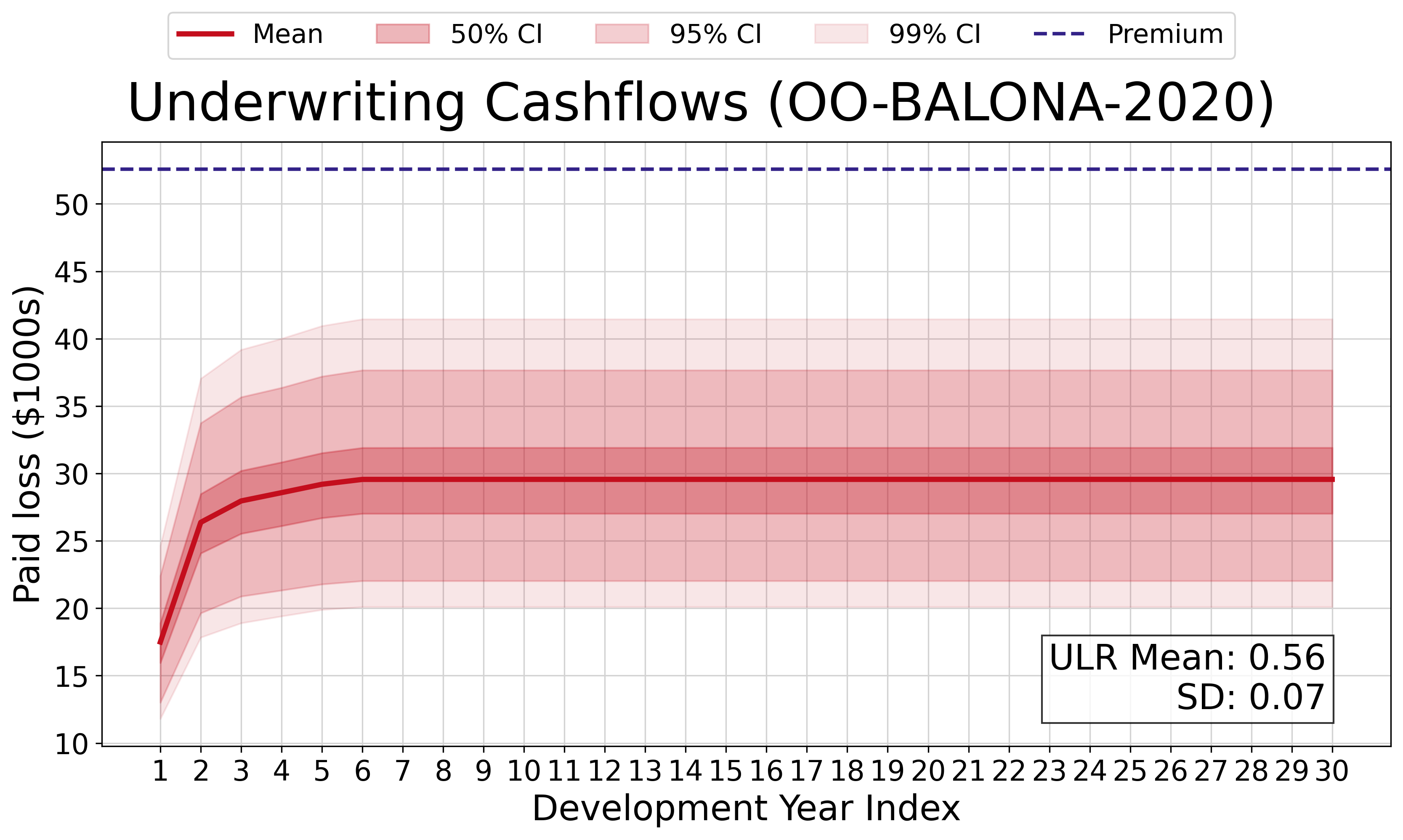}
    \caption{Predicted paid loss development for the one-year-ahead accident year for the Swiss liability triangle from \cite{balona2020}. Paid losses for future accident years represent the expected underwriting cashflows associated with the ILS deal. The shaded regions correspond to the 50\% and 95\% credible intervals for the predictions.}
    \label{fig:cashflows-OO-balona-2020}
\end{figure}

Figure \ref{fig:cashflows-OO-balona-2020} shows the predicted underwriting cashflows for the Swiss 
liability triangle. It is clear from Figure \ref{fig:cashflows-OO-balona-2020} that paid losses are 
expected to converge to the ultimate loss after approximately 6 years, with an expected loss ratio of 
56\% and a 97.5\%ile loss ratio of about 72\%. Given the relatively low expected ultimate loss ratio 
along with the development pattern, this program is a good candidate for securitization (see Section \ref{section:investment-cashflows}).

\section{Discussion}
The Bayesian workflow presented in this paper offers a comprehensive framework for 
approaching the first hurdle necessary to securitize casualty insurance risk -- obtaining 
accurate, well-calibrated ultimate loss predictions for future accident years. Our 
simulations and real-world applications demonstrate that the Bayesian framework is not 
only powerful enough to account for complexities underlying casualty ILS, but that it is 
also maximally transparent. Specifically, the use of Bayesian models requires us to 
encode each of our assumptions about loss development and forecasting dynamics into a 
mathematical expression that can be thoroughly inspected and critiqued. Cross-validation 
(i.e. backtesting) then shows us whether our model is expected to perform well enough for 
real-world applications. We believe these methods are crucial for ensuring that casualty 
ILS deals are priced in a data-driven way, which increases transparency and trust between 
insurers, captial providers, and analysts alike.  

There is still much work to be done in the casualty ILS space both on loss development 
and loss forecasting. For example, for loss development, we have found that the 
selection of which development lags should be used for training the body versus tail 
models (i.e., $\tau$ and $\rho$ in equations \ref{eq:chain-ladder} and \ref{eq:bondy}, 
respectively) can have a large influence on ultimate loss predictions. This finding 
inspired work by \cite{goold2024}, who showed that hidden markov models can be used to 
model how the loss development process switches from the body to tail state as 
development progresses. Despite showing similar or better performance on out-of-sample 
tests when compared to using traditional training window methods, both the hidden 
markov and traditional approach resulted in relatively poor calibration, emphasizing 
the need for further research on the core assumptions underlying common loss 
development models. 

Similarly, for forecasting models, how uncertainty is propegated from the loss 
development stage into the forecasting stage has strong implications for the forecasted 
ultimate losses. In the workflow we presented, we took an error-in-variables approach 
(see equestion \ref{eq:measure-error}). However, there are multiple other reasonable 
approaches to take, the most obvious being jointly modeling the loss development and 
forecasting stages. Historically, we have avoided doing so for practical reasons as 
mentioned in section \ref{section:measure-error}. However, we believe this could be a 
good area for future research -- the inconvenience of the joint model could be outweighed
by potential performance improvements. Further, ultimate loss ratios from year-to-year 
can be highly impacted by general market conditions \citep{berger1988}, leading to market 
cycles that have an observable impact on loss ratios. Models that can leverage market 
cycle information therefore have strong potential to improve ultimate loss forecasts. Our 
Bayesian framework is well suited to test if such models actually lead to meaningful 
improvements in accuracy (or calibration) over models that do not account for industry
dynamics. 

More generally, this is the first paper of its kind to introduce a fully worked-out 
Bayesian workflow to address casualty insurance loss modeling. In the same way that 
models used throughout actuarial science have been studied, extended, and critiqued, we 
believe that actuarial workflows should be formalized and rigorously studied in an 
effort to both improve and disseminate them. To support future research in this vein, 
we have made all of the scripts used throughout this paper available in a repository 
(\url{https://github.com/LedgerInvesting/bayesian-workflow-paper-2024}). In doing this 
work, it has also become clear to us that future research on ILS (and throughout 
actuarial science more generally) would greatly benefit from the availability of richer 
open datasets that can be used to test model assumptions. Standardized open datasets 
would allow for researchers across institutions to compare models against common 
benchmarks, accelerating model developent throughout actuarial science. The 
\cite{meyers2015} dataset is a great starting place, but any results based on it should 
be taken with a grain of salt given its size and lack of recency. Similarly, the Swiss liability 
triangle from \cite{balona2020} is a good candidate for testing model performance given its 18 year 
history. However, it is just a single triangle, and the most recent accident years available are from 
2015. Larger datasets with longer, more recent histories and more lines of business are needed to ensure 
that published backtest results are generalizable. Although many private organizations have access to 
such data (e.g., through purchase agreements with data aggregators), licensing restrictions prohibit 
open use of these datasets. The impetus then is on the research community to advocate for more data 
sharing and collaboration, which could ultimately lead to more robust and reliable models in the field.

\bibliographystyle{apa}
\bibliography{\src/references.bib}
\end{document}